\documentclass[11pt]{article}

\usepackage{graphicx}
\usepackage{authblk}
\usepackage{amsmath}
\usepackage[a4paper,margin=2.2cm]{geometry}
\usepackage{hyperref}
\usepackage{caption}
\usepackage{float}
\usepackage{booktabs}
\usepackage[numbers]{natbib}

\usepackage{array,tabularx,booktabs,siunitx,ragged2e}
\newcolumntype{Y}{>{\RaggedRight\arraybackslash}X} 

\usepackage{wrapfig}

\graphicspath{{./figures/}{./results/}}

\usepackage[caption=false,font=normalsize,labelfont=sf,textfont=sf]{subfig}

\newcommand{\mysubfigure}[3]{%
    \subfloat[#2]{\includegraphics[width=#1]{#3}}%
}
\usepackage[dvipsnames]{xcolor}
\definecolor{darkgreen}{rgb}{0.0, 0.5, 0.0}
\definecolor{brown}{rgb}{0.6, 0.3, 0.0}
\definecolor{darkbrown}{rgb}{0.4, 0.2, 0.0}

\usepackage{todonotes}

\DeclareGraphicsExtensions{.pdf,.jpeg,.png}

\usepackage{hyperref}
\usepackage{hyperxmp}
\hypersetup{
bookmarks=true,
bookmarksnumbered=true,            
hypertexnames=false,               
breaklinks=true,                   
linktoc=all,
colorlinks=true,
linkcolor=blue,
citecolor=blue,
urlcolor=blue,                     
pdfborder={0 0 112.0},              
hyperfootnotes=false
}                        

\title{A Distributed Acoustic Sensing Dataset for Vessel Detection and Localization in Submarine Cable Protection}

\author[1]{Erick Eduardo Ramirez-Torres}
\author[1,5]{Javier Macias-Guarasa}
\author[1]{Daniel Pizarro}
\author[2]{Javier Tejedor}
\author[1]{Sira Elena Palazuelos-Cagigas}
\author[1]{Pedro J. Vidal-Moreno}
\author[1]{María R. Fernández-Ruiz}
\author[1,3]{Sonia Martin-Lopez}
\author[1,3]{Miguel Gonzalez-Herraez}
\author[4]{Roel Vanthillo}

\affil[1]{Universidad de Alcal\'a, Departamento de Electr\'onica, Alcal\'a de Henares, Spain}
\affil[2]{Institute of Technology, Universidad San Pablo-CEU, CEU Universities, Urbanización Montepríncipe, 28668 Boadilla del Monte, Spain}
\affil[3]{Daza de Valdés Institute of Optics (IO-CSIC), Madrid, Spain}
\affil[4]{Marlinks, Leuven, Belgium}
\affil[5]{Corresponding author}

\date{}

\begin{document}
\maketitle

\begin{abstract}
  Recent incidents of accidental damage and suspected sabotage to submarine telecommunication and power cables, particularly in the Baltic Sea, have underscored their vulnerability and the need for continuous monitoring solutions. Distributed acoustic sensing (DAS) applied to submarine optical-fiber cables enables wide-area monitoring of underwater acoustic activity.

  We present the \texttt{Marlinks-NS} DAS dataset, comprising processed submarine DAS measurements and AIS-derived vessel information curated for cable-protection research. The dataset defines two machine-learning tasks (vessel detection and vessel-to-cable distance estimation) allowing reproducible research under realistic marine conditions.

  The dataset contains 74,771 labeled data instances from ten days of continuous recording along a $2,554\,m$ segment in a $28\,km$ buried fiber-optic cable in the North Sea. Each instance includes spectral-energy features from 250 sensing channels, together with anonymized distance measurements and metadata from AIS information. The released HDF5 data, documentation, processing description, and example code support reproducible development and evaluation of DAS-based vessel-monitoring methods for submarine cable protection.

\end{abstract}

\section{Background \& Summary}
\label{sec:background}

Submarine cables are vital for global communication and power transmission but remain vulnerable to accidental damage and sabotage, with recent severe examples in the Baltic Sea. These include, for example, the damage to the EstLink2 power cable~\cite{navalnews_estlink2_2024}, and the C-Lion1 communication cable~\cite{euromaidan_baltic_2025}, both suspected of sabotage and linked to geopolitical tensions.

Distributed Acoustic Sensing (DAS) technology allows repurposing of standard optical fibers into dense acoustic sensor arrays capable of capturing vibrations and strain over tens of kilometers. In the context of submarine cables, this capability enables continuous monitoring of marine activity using existing or dedicated fiber-optic infrastructure.
Unlike radar or satellite-based systems, DAS provides continuous, real-time coverage, is largely unaffected by lighting or weather conditions, and does not depend on cooperative identification systems such as the Automatic Identification System (AIS).

Publicly available datasets enabling the development of data-driven algorithms for DAS-based vessel detection are scarce. Existing data resources are often limited in scale, or lack properly labeled metadata that allow research in Artificial Intelligence/Machine Learning (AI/ML) strategies. This data scarcity is particularly severe in specialized environments like submarine cables. Initiatives such as PubDAS~\cite{pubDAS2023} (the first large-scale open-source repository for DAS data) have allowed accelerated research by providing multiple experiment datasets for benchmarking and algorithm development. However, the number of publicly available resources in submarine environments is very limited. Some relevant examples in this scenario are:
\begin{itemize}
  \item The DAShip dataset, by Huang et al.~\cite{Huang2025DAShip} (available at \cite{DAShip2025}).
  \item The MARS DAS experiment dataset by Cheng et al.~\cite{cheng2021utilizing} (available at~\cite{cheng2020mbari_das_data} with a \href{https://github.com/njlindsey/Photonic-seismology-in-Monterey-Bay-Dark-fiber1DAS-illuminates-offshore-faults-and-coastal-ocean}{repository at GitHub}).
  \item The MEUST DAS dataset by Sladen et al.~\cite{sladen2019distributed} (available at~\cite{sladen2019meust_km3net_das}).
  \item The HCMR+NESTOR+MEUST DAS datasets by Lior et al.~\cite{lior2021detection} (available at~\cite{lior2021detection_dataset}).
  \item The DAS4Microseism dataset, by Taweesintananon et al.~\cite{Taweesintananon2023} (available at~\cite{VPRD2H_2022}).
  \item The OOI Regional Cabled Array (RCA) Dataset, by Lipowsky et al.~\cite{shi2025multiplexed} (available at \cite{lipovsky2024rapid_das_ooi} with a \href{https://github.com/uwfiberlab/OOI_DAS_2024}{source code repository at GitHub}).
  \item The Valencia-IslaLink DAS dataset, by Spica and Gaite, which is distributed within PubDAS~\cite{pubDAS2023}.
  \item The Yellow River Delta Dataset, by Song et al.~\cite{song2024near} (available at~\cite{song_2024_dataset}).
\end{itemize}

\begin{table*}[t]
\centering
\footnotesize
\setlength{\tabcolsep}{3pt}     
\renewcommand{\arraystretch}{1.15}
\caption{Summary of publicly released DAS datasets cited in this work, including application domain, fiber length, dataset size, sampling frequency, gauge length and annotation availability.}
\label{tab:das_datasets}
\begin{tabularx}{\linewidth}{
  >{\RaggedRight\arraybackslash}p{2.8cm}
  >{\RaggedRight\arraybackslash}p{3.3cm}
  >{\RaggedRight\arraybackslash}p{2.0cm}
  >{\RaggedRight\arraybackslash}p{2.0cm}
  >{\RaggedRight\arraybackslash}p{1.8cm}
  Y
}
\hline
\textbf{Dataset} &
\textbf{Application domain} &
\textbf{Fiber length} &
\textbf{Dataset size} & \textbf{Sampling frequency and gauge length (GL)} &
\textbf{Annotation}\\
\hline
DAShip~\cite{Huang2025DAShip,DAShip2025} &
Wake ship detection and vessel speed/angle classification &
8.5\,km &
55875 ship passing events above the cable &
10\,Hz GL=4\,m&
Annotated using AIS guidance (for the three binary classification tasks)\\
\hline
MARS DAS experiment~\cite{cheng2021utilizing,cheng2020mbari_das_data} &
Ocean dynamics \& subsurface characterization &
51\,km &
4 days, 3.2TB &
250\,Hz GL=10\,m&
No labeling is provided \\
\hline
MEUST~\cite{sladen2019distributed,sladen2019meust_km3net_das} &
Seismic monitoring &
41.5\,km &
6 days &
1\,kHz GL=19.2\,m &
No labeling is provided \\
\hline
HCMR+ NESTOR+ MEUST~\cite{lior2021detection,lior2021detection_dataset} &
Seismic monitoring &
13.2\,km, 26.2\,km and 44.8\,km &
1 day (68 GB), 7 days (740 GB), 20 days (16 TB) &
Between 100\,Hz and 500\,Hz GL=19.2\,m/10\,m &
No labeling is provided \\
\hline
DAS4Microseism \cite{Taweesintananon2023,VPRD2H_2022} &
Ocean dynamics \& subsurface characterization &
120\,km &
44 days &
50\,Hz GL=8-16\,m&
No labeling is provided \\ 
\hline
OOI RCA~\cite{shi2025multiplexed,lipovsky2024rapid_das_ooi} &
Seismic monitoring &
~95\,km &
4 days, 2.4TB &
200\,Hz GL=40.852\,m &
No labeling is provided\\
\hline
Valencia~\cite{pubDAS2023} &
Submarine monitoring vibrations (telecommunication cable) &
$\sim$50\,km (first $\sim$10\,km on land) &
7 days, 3TB &
250\,Hz GL=30.4\,m&
No labeling is provided \\ 
\hline
Yellow River Delta~\cite{song2024near,song_2024_dataset} &
Nearshore ocean current monitoring &
10\,km &
25 days &
50\,Hz  GL=5\,m&
Provide labels for wind speed and tidal data \\
\hline
\end{tabularx}
\end{table*}

Table~\ref{tab:das_datasets} shows the relevant characteristics of each of them. Except for DAShip, all these datasets are oriented to applications in the seismic and ocean monitoring domain without explicit labeling information that can support experimental work in vessel detection and localization. DAShip is a large-scale, annotated dataset collected for the specific purpose of marine vessel detection. It comprises 55,875 ship-related event segments, captured on a submarine fiber-optic cable, $8.5\,km$ long, with a maximum depth of $13\,m$ along the cable deployment (according to GEBCO bathymetry information~\cite{gebco2024}).

It is oriented towards detecting vessel passages directly above the fiber-optic cable by applying image-classification algorithms to vessel-wake signatures. This formulation is less suited to early-warning cable-protection scenarios, in which vessels should ideally be detected well before reaching the cable. Another limitation is that the data have been downsampled to $10\,Hz$, which has two main issues: first, this reduced bandwidth limits the possibility of exploiting the higher frequency components known to be significant in vessel acoustic signatures~\cite{Rivet_2021,Thiem_2023}; and second, it will be more affected by low-frequency noise from temperature drift and the effect of sea waves and marine currents.

We can mention other properly labeled recent datasets which address event classification tasks, but in terrestrial domains, such as that by Tomasov et al.~\cite{tomasov2025comprehensive} that presented a fully labeled DAS dataset, including band-limited feature vectors. The fiber was $1{,}663\,m$ long, and was buried $1\,m$ below the surface around a rectangular area on an university campus. The selected events included walking, running, longboarding, and driving, as well as potential security-related events like fence climbing, fiber manipulation, and opening/closing of manholes.

The \texttt{Marlinks-NS DAS} dataset that we present in this paper addresses the lack of publicly accessible, well-annotated DAS data aimed at submarine cable protection applications, by defining two AI/ML tasks, one for vessel detection, and another for vessel-to-cable distance estimation. The dataset provides a large-scale, machine-learning-ready collection of preprocessed spectral features and vessel metadata derived from a ten-day continuous recording campaign on a submarine cable. It comprises $74{,}771$ data instances (feature vectors) computed at $250$ channels in a relevant fiber-optic sensing range, each containing energy values in $100$ logarithmically-spaced frequency bands, together with synchronized vessel distance labels, plus anonymized vessel-related information obtained from AIS (type, beam, and length). By releasing this curated dataset, we aim to promote transparency and reproducibility in AI/ML-based DAS research while enabling the scientific community to benchmark vessel detection and localization algorithms under realistic conditions.
Basic data-handling examples are archived with the dataset on Zenodo, while additional actively maintained reproducibility tools are available in the companion GitHub repository at \url{https://github.com/UAH-PSI/das-vessel-detection}.
\begin{figure}[H]
  \centering
  \includegraphics[width=\textwidth]{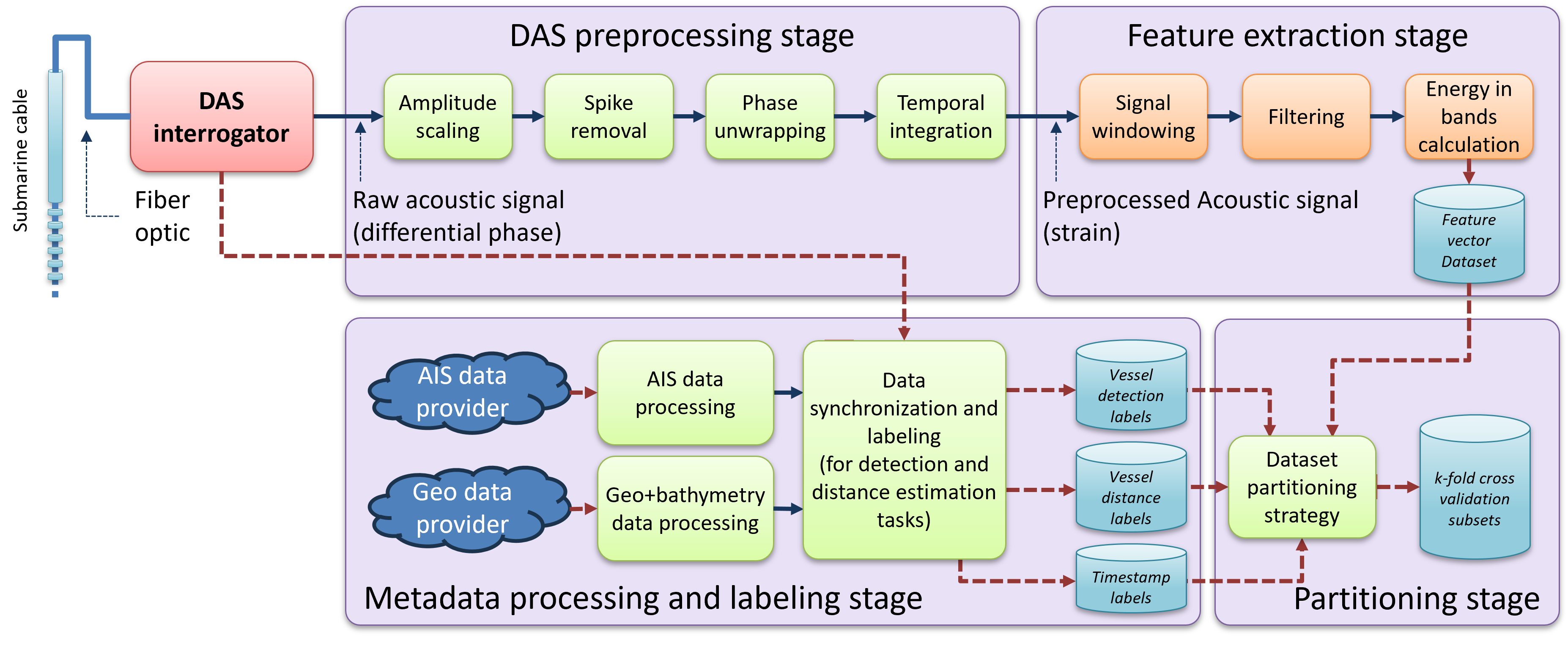}
  \caption{Dataset curation and processing pipeline, covering DAS acquisition and raw signal preprocessing; spectral feature extraction; metadata (AIS, geographical location and bathymetry) processing, synchronization and label generation; and cross-validation partitioning.}
  \label{fig:architecture}
\end{figure}

The DAS and AIS measurements underlying this dataset were previously analyzed in~\cite{RamirezTorres2026VesselDAS} to develop and evaluate machine-learning methods for vessel detection and vessel-to-cable distance estimation. That research article focused on algorithm design and application performance, including spatial and temporal aggregation, model comparisons, beamforming-based localization, vessel-dependent error analysis, and computational requirements. The present Data Descriptor provides dataset-centered documentation and reproducibility material that complements that application-centered study, rather than repeating its broader experimental analysis. Specifically, it provides a more detailed account of dataset generation and curation, the released feature representation, HDF5 organization, repository contents, metadata, known limitations and archived data-handling code, together with a new passage-level characterization of maritime-traffic diversity across the recommended day-wise folds, an expanded characterization of environmental variability, and baseline technical validation to support independent reuse.

\section{Methods}
\label{sec:methods}

The processing pipeline we applied for the generation of the dataset is shown in Fig.~\ref{fig:architecture}, to provide a full reference about the data curation process. All the relevant methods and modules are described in the following subsections.

\subsection{DAS interrogator: Acquisition Method and Setup}
\label{sec:acquisition-method}

The submarine fiber-optic cable was interrogated using a phase-sensitive Optical Time Domain Reflectometry ($\phi$-OTDR) system. This approach uses a narrow-linewidth, highly-coherent laser that emits short optical pulses into the fiber core, where microscopic refractive index variations act as Rayleigh scatterers. The backscattered signal from each pulse carries amplitude and phase information that varies with longitudinal deformation in the fiber, which can be related to vibrations and acoustic disturbances in the cable vicinity. By measuring phase differences between consecutive backscatter traces, dynamic strain variations can be reconstructed continuously along the entire length of the fiber.

\begin{figure*}[hbtp]
  \centering
  \mysubfigure{0.292\textwidth}{Regional map.}{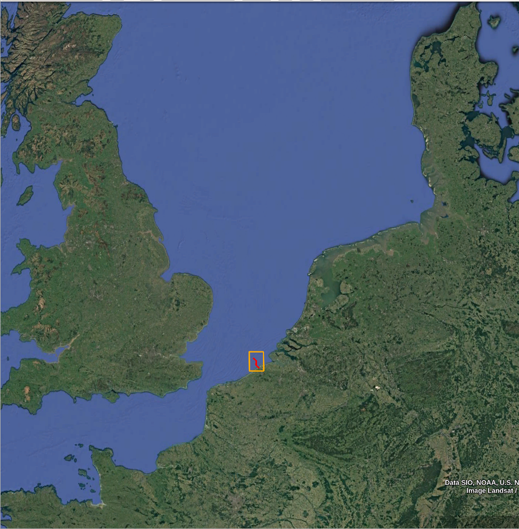}
  \label{fig:regional-map}
  ~
  \mysubfigure{0.290\textwidth}{Local map showing cable (red line).}{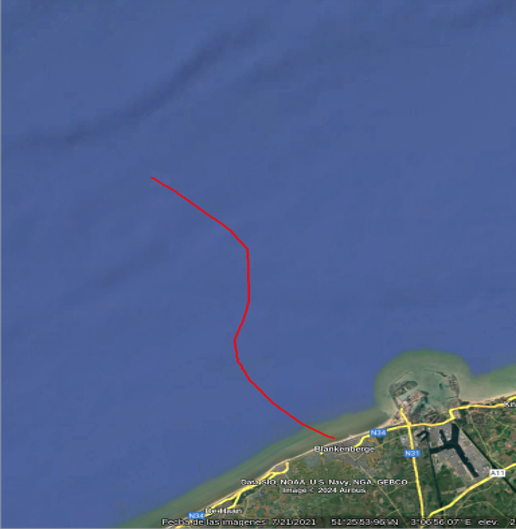}
  \label{fig:local-map}
  ~
  \mysubfigure{0.378\textwidth}{Local map showing cable (red line) and  bathymetry.}{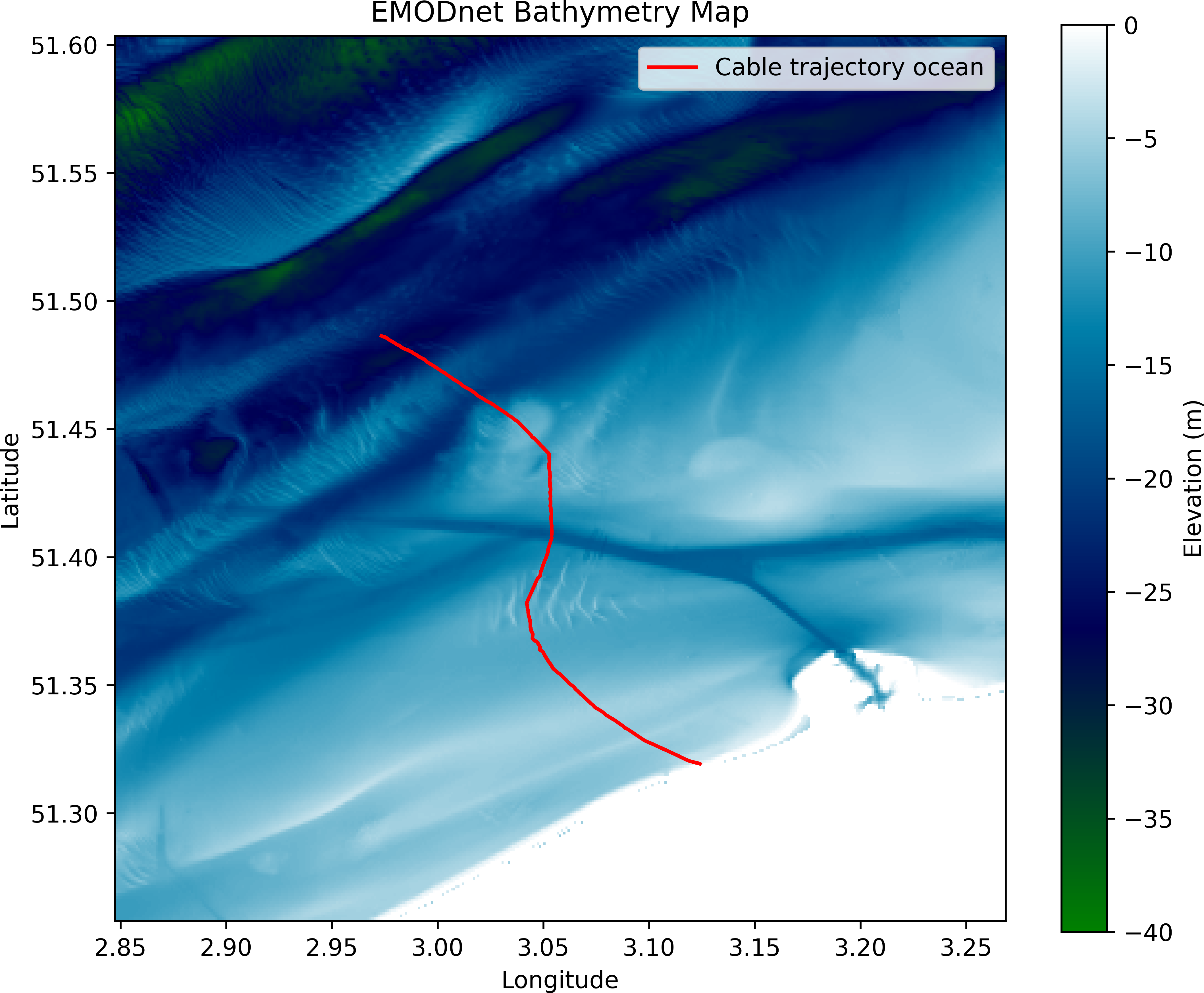}
  \label{fig:bathymetry-map}
  \caption{General location and bathymetry (cable location has been displaced for security considerations), from~\cite{RamirezTorres2026VesselDAS}.}
  \label{fig:geographical-location}
\end{figure*}

The DAS data were recorded during a ten-day acquisition campaign from 16 to 25 June 2023 using an Alcatel OptoDAS interrogator~\cite{OptoDAS} connected to a $28\,km$ ocean-bottom fiber-optic cable originally deployed for power cable monitoring. The cable lies buried between $1.4-7.2\,m$ below the seafloor, offshore Zeebrugge (Belgium) in the North Sea, as shown in Fig.~\ref{fig:geographical-location}. The interrogator is located off-shore and employs optical pulse-compression reflectometry, which enables distributed vibration sensing over tens of kilometers with meter-scale spatial sampling, while mitigating fading effects due to low-intensity points in the Rayleigh pattern \cite{gabai2016sensitivity, Zou_2015, Waagaard_2021}. It has a gauge length of $L=10.21\,m$, hence generating $2{,}774$ raw spatial channels. The raw differential phase-rate signals were sampled at $f_s=3{,}125\,Hz$.

\subsection{DAS Preprocessing}
\label{sec:signal preprocessing}

In this stage, the raw differential phase-rate time series generated by the DAS interrogator are converted to strain. This procedure is very closely related to the interrogator design and characteristics, so that the corresponding modules are mainly provided by the cable operator. The operations carried out are:

\begin{itemize}

  \item {Raw-signal scaling:} The integer samples delivered by the OptoDAS system were mapped to differential phase rate using the instrument scaling factor. Processing was performed separately for each channel along the time axis.

  \item {Impulse suppression:} Before integration, isolated spikes were identified using a channel-dependent local criterion. The threshold combined a 50-sample running mean with an offset of $30~\mathrm{rad/(m\,s)}$, and samples exceeding it in magnitude were replaced by zero. These parameters were selected empirically from separate DAS recordings outside the released campaign.

  \item {Temporal phase unwrapping:} Each channel was unwrapped independently in time to remove discontinuities of $2\pi$.

  \item {Strain recovery:} The cleaned and unwrapped phase-rate sequences were integrated over time and converted into strain, expressed in $\mathrm{m/m}$.
\end{itemize}

\begin{figure}[bthp]
  \centering
  \includegraphics[width=\textwidth]{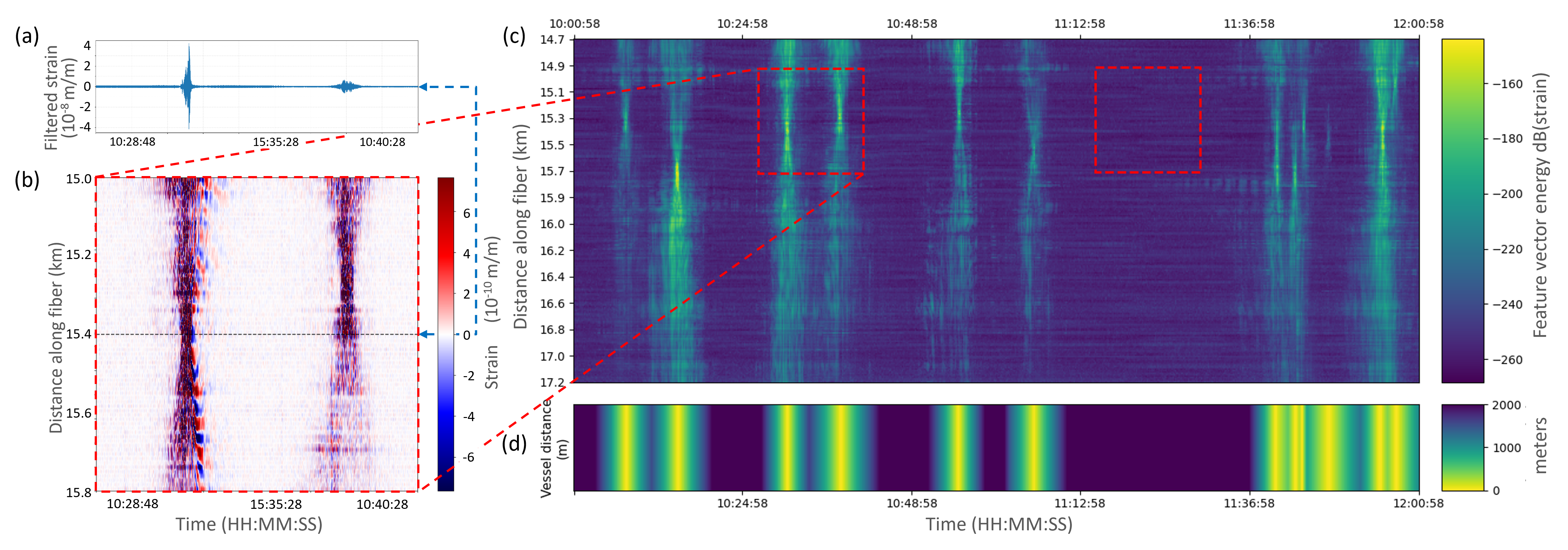}
  \caption{(a) Filtered strain signal when a vessel is near the cable sensed at $15.4\,km$ in the time interval of the left red dashed square in Fig.~\ref{fig:energy-dt-map}.c. (b) Filtered strain data as a function of time and distance along the cable corresponding to the left red dashed square in Fig.~\ref{fig:energy-dt-map}.c, time synchronized with Fig.~\ref{fig:energy-dt-map}.a. (c) Feature vector energy (in dBs) as a function of time and distance along the cable. (d) Distance from cable to nearest vessel (in meters), time-synchronized with Fig.~\ref{fig:energy-dt-map}.c.}
  \label{fig:energy-dt-map}
\end{figure}

Figs.~\ref{fig:strain-silence}.a+b and~\ref{fig:energy-dt-map}.a+b show two examples of the generated strain signals for scenarios in the absence of nearby vessels, and with a vessel nearby and, respectively.

\subsection{Metadata Processing: AIS Data Processing}
\label{sec:ais-data-processing}

The AIS data covering the same acquisition period, provided by the cable operator, included vessel identifiers (MMSI), timestamps, coordinates, course, and speed for 745 unique vessels with 64,417 position reports. A relevant issue when using AIS data is the unpredictable and typically low position update rates~\cite{emmens2021promisesAIS}. A detailed analysis on our DAS data showed that $88\%$ of vessels reported positions only every 1 to 3 minutes, which may imply a significant positional uncertainty. To overcome this issue, we applied a linear interpolation between consecutive reported positions at $1\,Hz$ position update rate, excluding periods in which vessels did not report for time intervals over 60 minutes. Because the AIS records were the only available source of vessel positioning, the accuracy of the interpolated trajectories could not be assessed against an independent reference. Linear interpolation was retained as a transparent baseline method. The resulting distance labels therefore remain subject to uncertainty arising from the AIS reporting interval and from vessel motion between consecutive reports. The application-based results reported in Section~\ref{sec:technical validation} indicate that these labels retain information relevant to the defined ML tasks, but they do not constitute a direct validation of interpolation accuracy. More sophisticated trajectory-reconstruction methods could be investigated in future work~\cite{guo2021improved}.


After the interpolation process, we had over $1,200,000$ AIS position entries, which we further enriched with web-scraped additional AIS metadata (namely, vessel type, length and beam) that can be useful in further studies or AI/ML tasks requiring information related to vessel type and size.

\subsection{Metadata Processing: Geographical+Bathymetry Data Processing}
\label{sec:geospatial-tagging}

In our DAS context, geospatial tagging refers to the process of assigning precise geographical coordinates, and depth or elevation to each sensed position along the fiber-optic cable. In DAS applications, accurate geospatial tagging is crucial because it relates the distributed acoustic measurements to their exact physical locations along the fiber. This enables spatially consistent signal interpretation, correlation with environmental and bathymetric data, and the development of position-dependent models for event localization and propagation analysis~\cite{holman2025}.

In our dataset, each DAS sensing position along the selected cable segment was assigned precise coordinates provided by the data owner, which resulted in a complete set of geospatial attributes (longitude, latitude, depth) for all the DAS channels, which provides a fundamental reference in our geographically-informed ML tasks using the \texttt{Marlinks-NS DAS} dataset.

Cable and vessel horizontal coordinates were represented in the WGS 84 geographic coordinate system (EPSG:4326), using owner-supplied cable elevations and an assumed vessel elevation of 0 m; coordinates were transformed to WGS 84 Earth- centered, Earth-fixed coordinates (EPSG:4978) using EcefKarney from PyGeodesy 23.1.9, and vessel-to-cable distances were computed as the minimum three-dimensional Euclidean chord distance between each vessel position and the interpolated cable sensing points.


\subsection{ML Tasks and Data Balance Considerations}
\label{sec:ml-tasks}

In ML applications, maintaining a balanced dataset helps ensure unbiased model training and fair performance evaluation. For classification tasks, balance in the number of samples per class prevents the model from favoring dominant categories. For regression tasks, an even distribution of target values enables the model to learn the underlying functional relationships for the full data range, rather than overfitting to the most densely sampled regions. However, achieving such balance is often difficult in real-world scenarios, where natural or operational processes generate highly uneven data distributions, which is usually the case in AI/ML-based DAS applications.

The \texttt{Marlinks-NS DAS} dataset has been designed to support two complementary ML tasks related to vessel monitoring for submarine cable protection. The first task, \emph{vessel detection}, is formulated as a binary classification problem that is oriented to discriminate whether there is a vessel closer than a given distance threshold to the cable or not. The second task, \emph{distance estimation}, is defined as a regression problem in which the model aims to predict the distance between the cable and the nearest vessel. In both cases, the required labeling information is derived from the synchronized AIS data.

\begin{figure}[hbtp]
  \centering
  \includegraphics[width=\columnwidth]{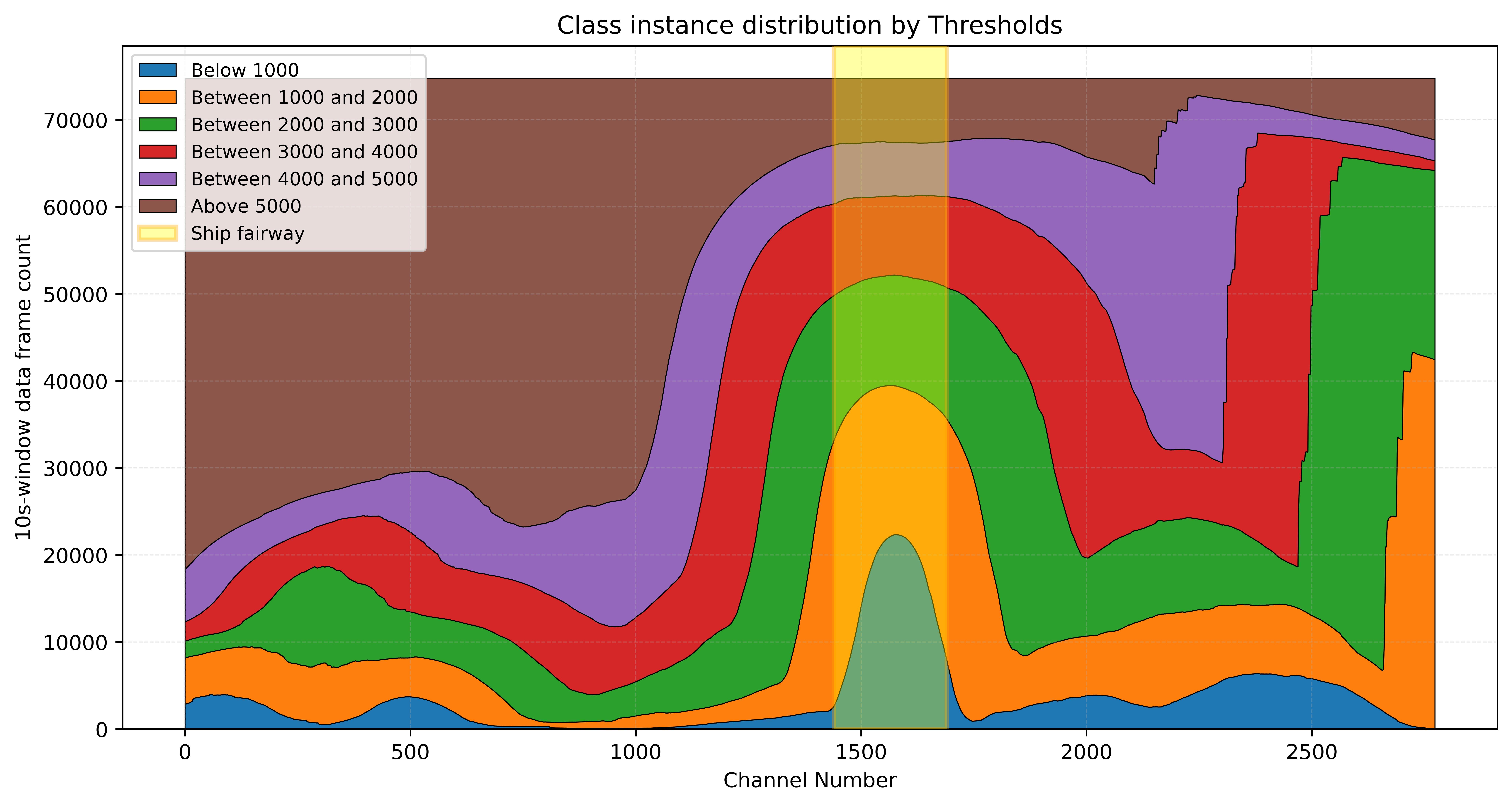}
  \caption{Number of data frames corresponding to vessels located at different distance ranges from the cable.}
  \label{fig:data-balance}
\end{figure}

To assess whether both tasks are representative, and evaluate their statistical balance, a detailed analysis of the vessel distance distribution along the monitored cable was carried out. Initial inspection of the ten-day continuous recording revealed strong non-uniform spatial distribution. Fig.~\ref{fig:data-balance} shows, as a function of the channel position along the cable length, the number of 10-second data frames that corresponds to a situation in which there is a vessel at different distance ranges. From this figure we can identify a cable section around $16\,km$ (channel 1565, roughly in the middle of the vertical yellow transparent band in the figure) that exhibits a relevant increase in the number of available data frames with vessels closer to the cable than, for example, $1{,}000\,m$ or $2{,}000\,m$, which can be candidate thresholds for the vessel detection task. As an example of the rationale for defining a ``reasonable threshold'', a Spanish company monitoring submarine cables with AIS, sets a $1{,}000\,m$ distance threshold to trigger enhanced vessel tracking to assess potential threats to the cable.

This region corresponds to the location of the cross between the submarine cable and a dredged fairway (approximately $20\,m$ deep) providing harbor access, which is consistent with the higher density of vessels shown in Fig.~\ref{fig:data-balance}. Therefore, on the basis of this spatial distribution, we selected a contiguous cable range of $2{,}553\,m$,  corresponding to a subset of $N_{channels}=250$ equally-spaced spatial channels located between $14{,}702\,m$ and $17{,}255\,m$ from the interrogator. This segment exhibits a clear unbalance between the ``vessel-nearby'' and ``no-vessel-nearby'' classes (e.g. $22{,}347$ vs. $52{,}424$ for the $1{,}000\,m$ distance threshold), but provides a practical compromise between the availability of vessel-nearby observations and the preservation of a contiguous sensing range. 

\begin{figure*}[hbtp]
  \centering
  \includegraphics[width=\columnwidth]{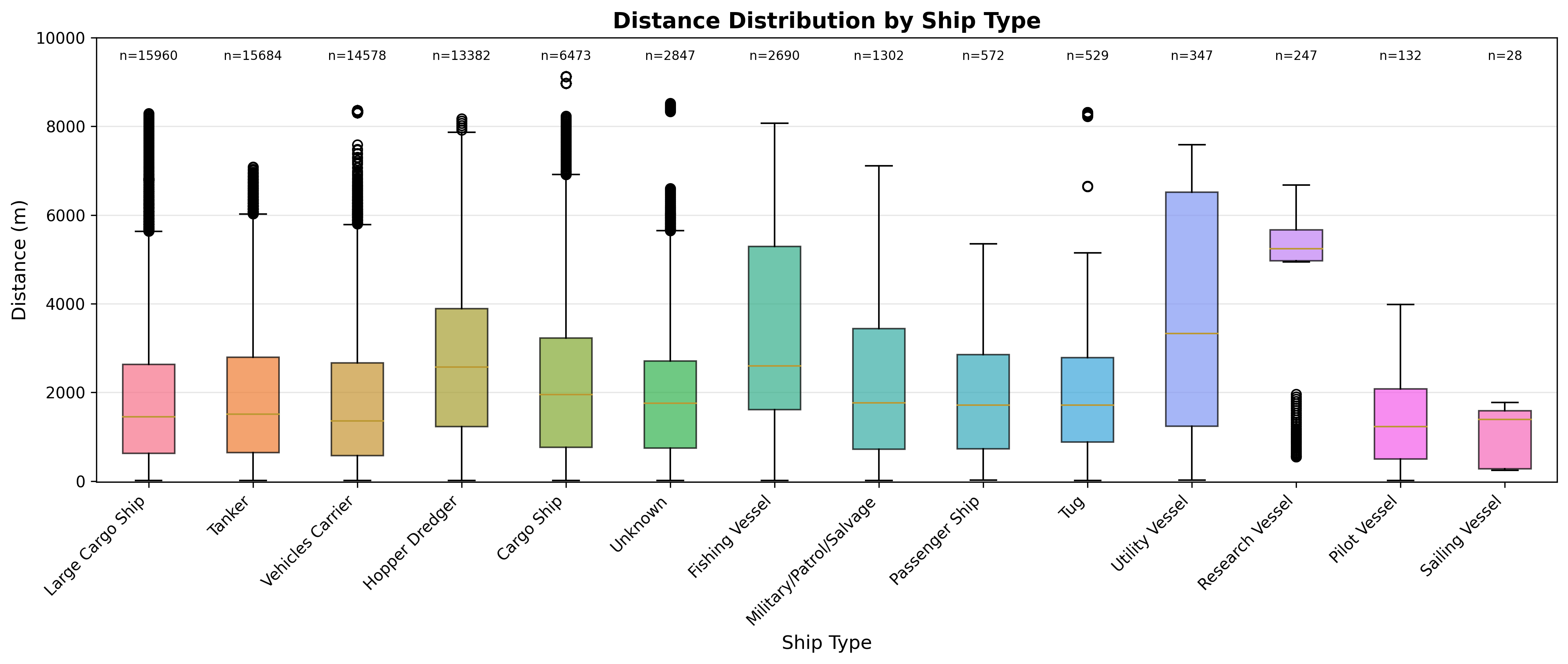}
  \caption{Boxplot of data frames available as a function of the \textit{minimum} distance of \textit{any} vessel to the cable, per vessel type.}
  \label{fig:piano-distance-by-type}
\end{figure*}

Finally, to give an idea on the distribution of vessel types, Fig.~\ref{fig:piano-distance-by-type} shows the distribution boxplots of minimum distances by vessel types, in which the top numbers ($n=\dots$) state the number of examples of each vessel type in the dataset.

\begin{figure}[htbp]
  \centering
  \includegraphics[width=\textwidth]{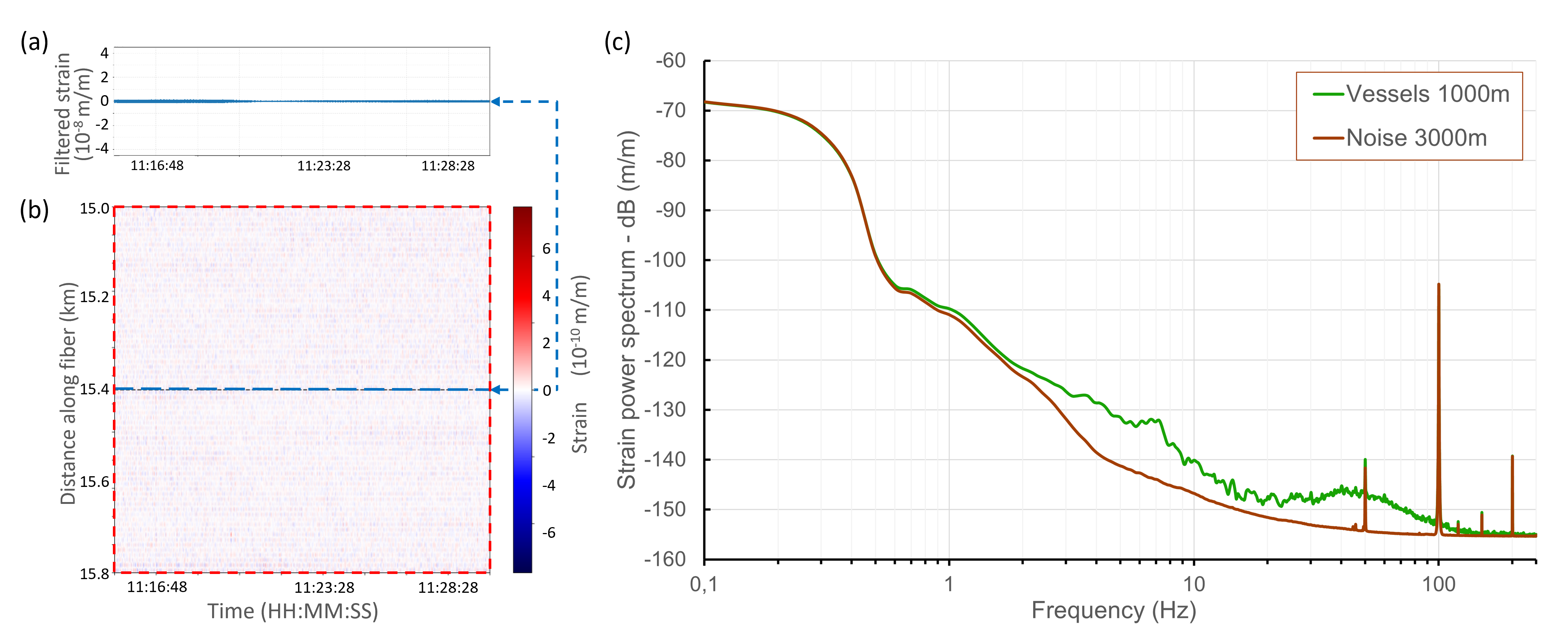}
  \caption{(a) Filtered strain signal when no vessel is nearby the cable sensed at $15.4\,km$ in the time interval of the right red dashed square in Fig.~\ref{fig:energy-dt-map}.c. (b) Filtered strain data as a function of time and distance corresponding to the right red dashed square in Fig.~\ref{fig:energy-dt-map}.c, time synchronized with Fig.~\ref{fig:strain-silence}.a. (c) Long term averaged strain power spectrum plots (expressed in dBs) for strain measurements for vessels closer than $1{,}000\,m$ to the cable (\textcolor{darkgreen}{\textrm{Vessels 1000m}} trace), and no vessels closer than $3{,}000\,m$ (\textcolor{darkbrown}{\textrm{Noise 3000m}} trace).}
  \label{fig:strain-silence}
\end{figure}

\subsection{Feature Extraction}
\label{sec:feature extraction}

To generate the dataset, we adopted a data-driven methodology to decide on the feature vector composition, informed by the available AIS data. The main idea was to compare long-term averaged strain spectra corresponding to vessel and background-noise conditions and identify frequency regions containing potentially discriminative information. Fig.~\ref{fig:strain-silence}.c shows the long-term averaged strain spectra (in dB) for the case of vessels closer than $1{,}000\,m$ to the cable (\textcolor{darkgreen}{\textrm{Vessels 1000m}} trace) and for the case of no vessels being closer than $3{,}000\,m$ (\textcolor{brown}{\textrm{Noise 3000m}} trace).

Based on this analysis, the lower cutoff was conservatively set to $4\,Hz$ to reduce the contribution of slow components associated with temperature variations, surface-wave loading, and other slowly varying environmental effects. Therefore, $4\,Hz$ should be understood as a processing choice intended to improve the robustness of the released representation, rather than as a precise physical boundary for vessel-generated signals.

The upper cutoff was selected considering both the observed spectra and the spatial response of the DAS acquisition configuration. Assuming that the gauge length was equal to the channel spacing, $L=\Delta x=10.21\,m$, and using the apparent propagation velocity estimated for the dominant propagation path, $c_{\mathrm{app}}\simeq1{,}750\,m/s$~\cite{RamirezTorres2026VesselDAS}, the nominal along-fiber spatial-aliasing limit is
$f_{\mathrm{alias}}=\nobreak c_{\mathrm{app}}/(2\Delta x)\simeq85.7\,Hz$.
Independently, the spatial averaging over the gauge length progressively attenuates the DAS response, with its first theoretical notch occurring at
$f_{\mathrm{notch}}=c_{\mathrm{app}}/L\simeq171.4\,Hz$
under the same propagation assumption.

The spatial-aliasing limit does not constitute a sharp temporal-frequency cutoff: components above it can still be measured at individual channels, although their spatial variation may be aliased, while gauge-length averaging increasingly attenuates them as the frequency approaches the first notch. Since the empirical spectra continued to show vessel-to-background differences close to $100\,Hz$, this frequency was retained as the upper limit of the initial analysis range, provided the very small amplitude of components at higher frequencies.

Additionally, we discovered strong narrowband spectral components around $100\,Hz$, with harmonics at its integer multiples and a subharmonic around $50\,Hz$, probably due to interference generated by rotating mechanical systems or electrical noise in the interrogator environment. Since these components were consistently present under both vessel and background-noise conditions, they were considered non-discriminative site- or instrument-related noise, and the bands associated with the nominal $49$--$51\,Hz$ and $98$--$100\,Hz$ intervals were excluded from the feature representation.

The strain signal for each channel was then segmented into non-overlapping windows of length $T_\omega=10\,s$, to which we apply a Blackman window to reduce spectral leakage. At the sampling frequency $f_s=3{,}125\,Hz$, each window contains $N_{\mathrm{FFT}}=31{,}250$ samples, resulting in a nominal FFT-bin spacing of
$\Delta f=f_s/N_{\mathrm{FFT}}=1/T_\omega=0.1\,Hz$.
The window length was selected as a compromise between temporal granularity and frequency-domain representation: it provides sufficiently fine spectral sampling in the low-frequency region while generating one feature vector every $10\,s$, allowing the temporal evolution of vessel-induced signals to be reasonably preserved. The FFT coefficients within the initial $4$--$100\,Hz$ analysis range were then retained for the subsequent band-energy calculation, with the noise-related intervals excluded during the frequency-band definition described below.

The feature vector is then generated from the FFT values, so that each 10-second window is represented by $N_{bands}=100$ logarithmically spaced band-energy features. To construct these bands, we first generated $N_0=104$ contiguous frequency intervals over the initial $4$--$100\,Hz$ analysis range, using the logarithmically spaced edges described by
\begin{equation}
  \label{eq:fi}
f_i=f_{\min}
\left(\frac{f_{\max}}{f_{\min}}\right)^{i/N_0},
\qquad i=0,1,\dots,N_0,
\end{equation}
where $f_{\min}=4\,Hz$ and $f_{\max}=100\,Hz$. $N_0=104$ is used so that the final number of frequency bands is  $N_{bands}=100$, given that bands overlapping the nominal $49$--$51\,Hz$ interference interval were subsequently discarded, together with the highest-frequency band overlapping the nominal $98$--$100\,Hz$ interval. This geometric construction provides a constant ratio between consecutive frequency edges and, therefore, an approximately constant fractional bandwidth. Consequently, the bands are narrower in absolute frequency at the lower end of the spectrum, providing a finer representation of the low-frequency region, and become progressively wider at higher frequencies. The exact limits of all the retained frequency bands are provided in the \texttt{fbands.csv} file, available in the archived Zenodo release and in the \href{https://github.com/UAH-PSI/das-vessel-detection/blob/main/data/fbands.csv}{companion GitHub repository}.

The mathematical details of the feature extraction calculation are as follows. We assume that $x_c^{(n)}[m]$ denotes the discrete version of the $n^{th}$ Blackman-windowed strain signal corresponding to channel $c$ and starting at $n\cdot T_\omega$ seconds, with $n=0,1,\dots,N_{windows}-1$ and $c=0,1,\dots,N_{channels}-1$, where $N_{windows}$ and $N_{channels}$ denote the number of available temporal windows and sensing channels, respectively. The sample index is $m=0,1,\dots,N_{\mathrm{FFT}}-1$. We then define $X_c^{(n)}[k]$ as the one-sided $N_{\mathrm{FFT}}$-point FFT of $x_c^{(n)}[m]$, with $N_{\mathrm{FFT}}=31{,}250$ and $k=0,1,\dots,N_{\mathrm{FFT}}/2$.

Each spectral band $b = 0, 1, \dots, N_{bands}-1$ is defined by its FFT-bin limits $\big[k_b^{-},\, k_b^{+}\big]$, where $0 \le k_b^{-} \le k_b^{+} < N_{\mathrm{FFT}}$ (corresponding to the frequency band limits defined in Eq.~\eqref{eq:fi}).

From these definitions, the spectral energy of band $b$ for window $n$ of channel $c$ is computed as:
\begin{equation}
E_c^{(n)}[b] = \sum_{k = k_b^{-}}^{k_b^{+}} \big| X_c^{(n)}[k] \big|^2.
\end{equation}

Finally, the feature vector for $x_c^{(n)}[m]$ will be $\mathcal{E}_c^{(n)} = \left[ E_c^{(n)}[0], E_c^{(n)}[1], \dots, E_c^{(n)}[N_{bands}-1] \right]$.
For each temporal window, the feature vectors from all channels form an $N_{channels}\times N_{bands}$ feature matrix. Stacking these matrices over time produces a tensor of dimensions
$N_{windows}\times N_{channels}\times N_{bands}$, which is stored in the \texttt{X} dataset of the released HDF5 file.

\subsection{Data Synchronization and Labeling}
\label{sec:ais data and label generation}

Time synchronization between the DAS signals and the interpolated AIS information is carried out as a required step before data labeling is carried out. The interpolated AIS positions are then used to continuously compute the shortest distance between each vessel and every DAS sensing point. These distance values are considered within each 10-second strain signal window to determine the minimum vessel-to-cable distance used as the primary label for the ML tasks. This synchronization process ensures that all DAS feature vectors are associated with consistent, temporally-aligned vessel metadata.

The closest distance continuous variable supports regression tasks directly as \texttt{Vessel distance labels}, or can be converted into \texttt{vessel detection labels} for the classification task, by simply applying user-defined distance thresholds.

\section{Data Records}
\label{sec:data records}

The dataset and its accompanying resources are publicly available on Zenodo~\cite{ramirez2024dasvesseldataset}. The primary data product is the approximately 14-GB HDF5 file \texttt{dataset\_sensor\_range\_1440\_1690\_0.h5}. The Zenodo deposit also includes dataset documentation in Markdown and PDF formats (\texttt{README.md} and \texttt{README.pdf}), supporting metadata in \texttt{misc.zip}, and simple Python data-handling examples in \texttt{src.zip}. The structure of the primary HDF5 data file is summarized in Table~\ref{tab:dataset-content} and described below:
\begin{itemize}
  \item \texttt{X}: A 3D NumPy array of shape {($N_{windows}$, $N_{channels}$, $N_{bands}$)}, containing the feature vectors in squared strain units $(m/m)^2$, captured by the sensors at each timestamp.
  \begin{itemize}
    \item \textbf{{$N_{windows}=74{,}771$}}: number of non-overlapped 10-second signal windows analyzed along the full recording period.
    \item \textbf{{$N_{channels}=250$}}: number of spatial channels (sensor positions along the selected fiber segment).
    \item \textbf{{$N_{bands}=100$}}: number of energy-band features per channel.
  \end{itemize}

  \item \texttt{y}: A 1D NumPy array of shape {($N_{windows}$)} containing the distance (in meters) to the closest vessel during each 10-second window. This continuous variable supports regression tasks directly, or can be converted into classification labels by simply applying user-defined distance thresholds.

  \item \texttt{datetimes}: An array of strings in the \texttt{HDF5} file that records the UTC timestamp (formatted as \texttt{\%Y-\%m-\%d \%H:\%M:\%S\%z} following Python’s \texttt{strftime} format-code conventions) corresponding to each 10-second window, with shape {($N_{windows}$)}.

  \item \texttt{ship\_info}: A group in the \texttt{HDF5} file containing AIS metadata of the vessel used to generate the \texttt{y} label:
        \begin{itemize}
          \item Vessel type, as a string value.
          \item Vessel length, in meters.
          \item Vessel beam, in meters.
        \end{itemize}
\end{itemize}

\begin{quote}
\textbf{Note on Noisy Sensors}

Three sensors (indices 59, 60, and 61 in the \texttt{X} array) consistently exhibited high noise levels and have been forced to zero in the raw data. We recommend excluding these channels from feature matrices prior to training classification and regression models.
\end{quote}

\begin{table}[H]
  \centering
  \caption{Summary of released dataset contents.}
  \label{tab:dataset-content}
  \begin{tabularx}{\linewidth}{@{}lXl@{}}\toprule
  \textbf{Component} & \textbf{Description} & \textbf{Dimensions} \\ \midrule
  \texttt{X} & 3D array of energy-band features per channel & $74{,}771 \times 250 \times 100$ \\
  \texttt{y} & Closest distance from the cable to any vessel [m] & $74{,}771$ \\
  \texttt{datetimes} & UTC timestamps (formatted as \texttt{\%Y-\%m-\%d \%H:\%M:\%S\%z} following Python’s \texttt{strftime} format-code conventions) for each 10-s window & $74{,}771$ \\
    \texttt{ship\_info} & AIS metadata for the vessel used to generate \texttt{y} labels &  \\
    & Vessel type & $74{,}771$ \\
    & Vessel length [m] & $74{,}771$ \\
    & Vessel beam [m] & $74{,}771$ \\
  \bottomrule
  \end{tabularx}
\end{table}

The public deposit does not include the original raw DAS recordings, complete AIS records, MMSI identifiers, nor the precise geographical route of the cable because of data-owner, confidentiality and critical-infrastructure restrictions.

\section{Technical Validation}
\label{sec:technical validation}

The technical validation was designed to assess complementary aspects of the released dataset and its recommended evaluation protocol. We first define the baseline machine-learning framework and the ten-fold day-wise partitioning used throughout the validation. We then examine maritime-traffic diversity and environmental variability across these folds to assess whether they represent distinct physical vessel passages and heterogeneous acquisition conditions. Finally, we report baseline classification and regression results to verify that the released feature representation and AIS-derived labels retain information relevant to the two defined machine-learning tasks. These analyses are intended to support dataset quality and reuse rather than to provide an extensive comparison of modeling approaches, which is reported separately in~\cite{RamirezTorres2026VesselDAS}.

\subsection{Machine Learning Framework and Day-Wise Evaluation Protocol}
\label{sec:machine learning framework}

We carried out the initial technical validation by applying an XGBoost ML algorithm to the two ML tasks defined in Section~\ref{sec:ml-tasks}.

The objective functions in the baseline XGBoost models are binary cross-entropy for classification and mean squared error for regression. We used gradient boosted trees, a learning rate $\eta = 0.05$, and a maximum tree depth of 10, with 500 boosting rounds.

Our experimental approach is based on a rigorous $k$-fold cross-validation strategy, in which the dataset is divided into $k$ equal parts (folds). A model is trained on $k-1$ folds and tested on the remaining one, repeating this process $k$ times so that each fold serves once as test data. The final performance metric is the average of the $k$ test results, giving a more reliable estimate of the model's performance, while keeping a strict separation between training and testing data. In our case, each fold corresponds to 1 full day of data, so that $k=10$ and the cross-validation is day-wise based. The available source code at the \href{https://github.com/UAH-PSI/das-vessel-detection}{companion GitHub repository} also promotes the use of the cross-validation strategy.

The vessel detection task was set with a distance threshold of $1{,}000\,m$, and the vessel-to-cable distance estimation task evaluated two cases: unrestricted distance estimation, and distance estimation for vessels closer than $1{,}000\,m$ to the cable.

We exploited spatial redundancy in the DAS channels by averaging the feature vectors from $N_C$ contiguous channels in the monitored fiber-optic segment, which is given as an input to the XGBoost algorithm.

\subsection{Maritime Traffic Diversity Across the Day-Wise Folds}
\label{sec:marit-traff-divers}

\begin{figure*}[t]
    \centering
    \begin{minipage}[t]{0.495\textwidth}
        \raggedright{\footnotesize (a)}\par\vspace{0.2em}
        \centering
        \includegraphics[width=\linewidth]{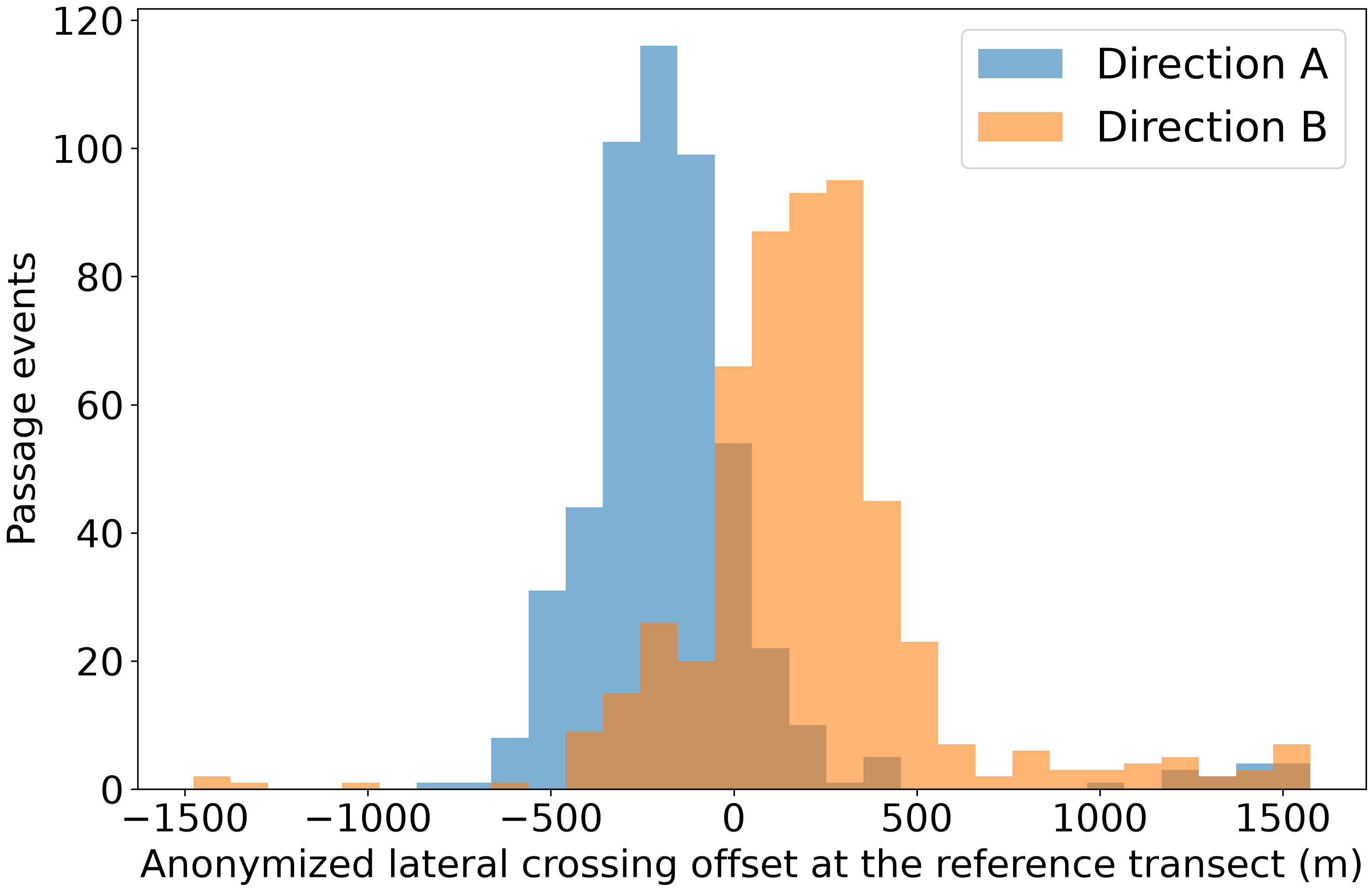}
    \end{minipage}%
    \hfill
    \begin{minipage}[t]{0.495\textwidth}
        \raggedright{\footnotesize (b)}\par\vspace{0.2em}
        \centering
        \includegraphics[width=\linewidth]{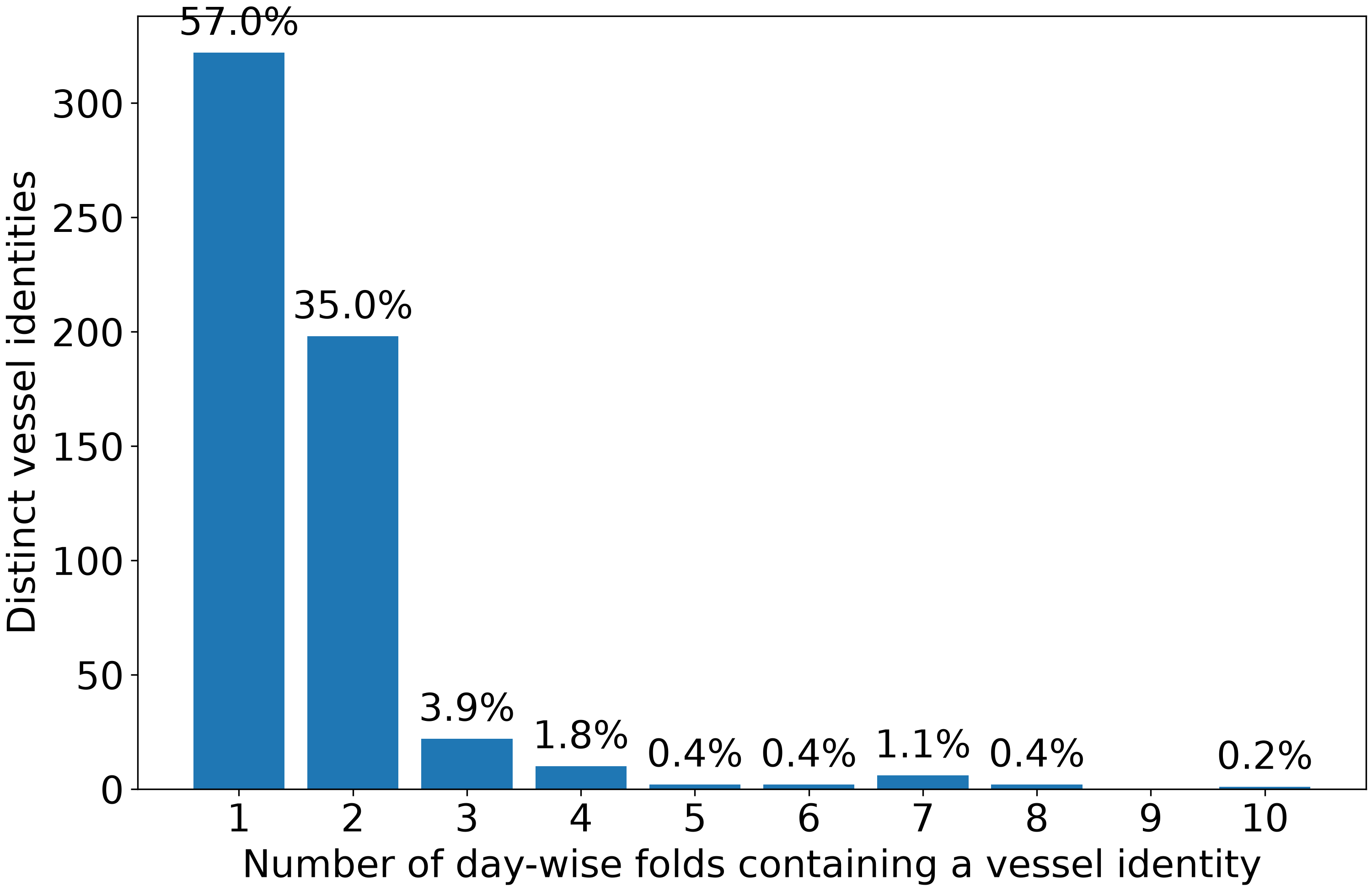}
    \end{minipage}
    \caption{Directional traffic structure and recurrence of vessel identities across the day-wise folds: (a) Distribution of the lateral crossing positions of the 1,033 reconstructed passage events, separated according to the two dominant travel directions. 
      (b) Distribution of the 565 unique vessel identities according to the number of day-wise folds in which each MMSI was observed.
    }
    \label{fig:traffic-characterization}
\end{figure*}

To characterize the local maritime traffic and assess the extent to which the day-wise folds represent distinct physical vessel passages within the monitored area rather than repeated observations of the same events, we analyzed vessel activity within a cable-adjacent region defined by expanding the bounding box of the monitored cable segment by $1\,km$ in all directions. This analysis comprised $4,290$ original AIS observations (without interpolation), representing $565$ unique MMSIs and $1,033$ reconstructed physical passage events\footnote{The original MMSI identifiers were used only for this post hoc characterization of vessel recurrence and are not included in the released feature representation or provided as inputs to the machine-learning models.}. The complete daily folds contained between 97 and 122 passages. 
Traffic was approximately balanced between two opposite travel directions, with 507 and 526 passages, respectively. Passage directions were strongly bimodal: $95.4\%$ of the reconstructed passages had AIS course-over-ground headings within two \(30^{\circ}\)-wide angular sectors centred on nearly opposite dominant travel directions. At a fixed reference transect approximately perpendicular to the dominant traffic direction, the corresponding directional traffic streams had median crossing positions separated by approximately $390\,m$, consistent with direction-dependent traffic lanes whose lateral distributions partially overlap as shown in Fig.~\ref{fig:traffic-characterization}.a.

Most vessel identities were transient during the observation campaign, as shown by their distribution across the day-wise folds in Fig.~\ref{fig:traffic-characterization}.b. Of the 565 MMSIs, 322 occurred on one day and 198 on exactly two days; thus, $92.0\%$ were present in no more than two daily folds. Among the 198 identities appearing on exactly two days, 192 generated only one passage on each day. For 189 of the 192 cases, the second passage occurred in the opposite traffic direction. These paired passages were separated by a median of $40.4\,h$ and differed in lateral crossing position by a median of approximately $311\,m$ (see distribution in Fig.~\ref{fig:traffic-characterization}.a). The identity recurrence observed between folds therefore primarily represents distinct outward and return passages rather than repeated measurements of equivalent traffic events.

Only four of the 1,033 reconstructed passages crossed a daily fold boundary (over midnight), corresponding to $0.39\%$ of all passage events and 16 of the 4,290 AIS observations. Among the test-fold MMSI occurrences for which the same MMSI also occurred in the training folds, $1.35\%$ corresponded to the same continuous physical passage. After excluding these boundary events, the minimum temporal separation from a same-identity training passage had a median of $36.7\,h$, and $81.8\%$ of the cases were separated by more than $24\,h$.

These results indicate that the day-wise partitioning is effectively passage-disjoint and predominantly evaluates new physical traffic events under realistic maritime conditions. Although vessel identities naturally recur in an operational monitoring region, such recurrence rarely corresponds to duplication of the same passage or adjacent DAS observations. The resulting folds therefore provide a robust and realistic basis for model validation, with negligible passage-level overlap and substantial variation in vessel identities, passage timing, travel direction, and crossing geometry.

\subsection{Environmental Variability Across the Day-Wise Folds}
\label{sec:environmental_variability}

\begin{figure}[htbp]
    \centering
    \includegraphics[width=.8\textwidth]{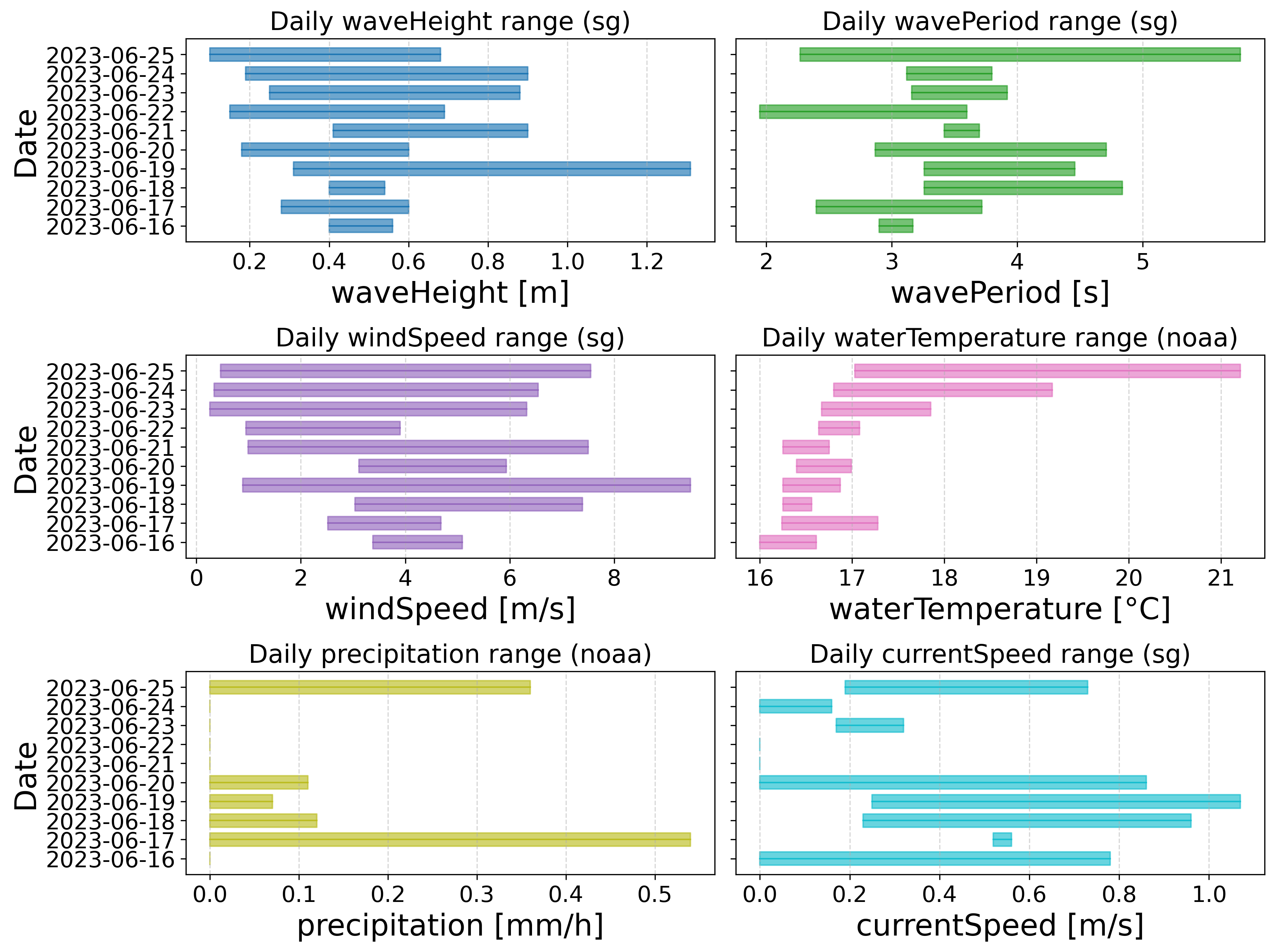}
    \caption{Daily minimum-to-maximum ranges of six environmental variables during the ten-day acquisition campaign.}
    \label{fig:meteorological_ranges}
  \end{figure}

To complement the maritime-traffic characterization, we examined the environmental conditions covered by the ten day-wise folds. Historical marine and meteorological data were retrieved through the Stormglass API at several query locations covering the monitored region. The variables considered were wave height, wave period, wind speed, water temperature, precipitation, and current speed. Fig.~\ref{fig:meteorological_ranges} shows the daily ranges at the closest location to the monitored region. Variables identified by the \texttt{sg} source correspond to the Stormglass aggregate source, whereas those identified by \texttt{noaa} correspond to NOAA-source values returned through the same API.

Under the applicable data-use conditions, the original time-resolved StormGlass API responses cannot be publicly redistributed. We therefore release the derived file \texttt{meteorological\_daily\_summary.csv}, which contains, for each UTC day, the minimum and maximum values of the six variables shown in Fig.~\ref{fig:meteorological_ranges}. These summaries are intended to document the environmental conditions represented in each fold, rather than to provide a replacement for the original meteorological data feed.

The daily ranges show substantial variation during the acquisition campaign. Across the ten days, wave height ranged from $0.10$ to $1.31\,m$, wave period from $1.95$ to $5.78\,s$, wind speed from $0.26$ to $9.46\,m/s$, water temperature from $16$ to $21.21^\circ$ C, precipitation from $0$ to $0.54\,mm/h$, and current speed from $0$ to $1.07\,m/s$. Both the positions and widths of the daily intervals varied across folds. For example, the largest wave-height and wind-speed values occurred on 19 June, whereas the highest water temperature and longest wave period occurred on 25 June. Precipitation was absent during most daily intervals but reached measurable values on a subset of days.

Consistent with this descriptive characterization, we carried out a statistical evaluation to assess whether the six meteorological variables differed significantly across the ten day-wise folds. Kruskal--Wallis tests, with Holm correction across the six variables, indicated significant day-to-day differences in wave height ($H(9)=113.99$, $p_{\mathrm{Holm}}=6.79\times10^{-20}$, $\varepsilon^2=0.458$), wave period ($H(9)=124.92$, $p_{\mathrm{Holm}}=5.25\times10^{-22}$, $\varepsilon^2=0.506$), wind speed ($H(9)=60.39$, $p_{\mathrm{Holm}}=1.13\times10^{-9}$, $\varepsilon^2=0.224$), water temperature ($H(9)=210.12$, $p_{\mathrm{Holm}}=1.50\times10^{-39}$, $\varepsilon^2=0.878$), precipitation ($H(9)=74.33$, $p_{\mathrm{Holm}}=4.29\times10^{-12}$, $\varepsilon^2=0.285$), and current speed ($H(9)=178.36$, $p_{\mathrm{Holm}}=5.57\times10^{-33}$, $\varepsilon^2=0.740$). A PERMANOVA performed on the standardized values of the six variables also detected significant multivariate differences in the combined environmental conditions among days ($\mathrm{pseudo}\text{-}F=24.01$, $R^2=0.485$, $p=1.0\times10^{-4}$; 9,999 permutations). Taken together, these results show that the day-wise folds were acquired under measurably different marine and meteorological conditions rather than under a nearly stationary environmental state.

This analysis is intended to characterize the environmental diversity of the released dataset and the recommended validation folds. It should not be interpreted as demonstrating that model performance is independent of meteorological conditions, nor as establishing representativeness across the full seasonal range of the deployment region.

\subsection{Baseline Validation Results}
\label{sec:results}

The evaluation metrics used depended on the ML task:

\begin{itemize}
  \item For the classification task (vessel detection): Accuracy, and global and class-wise $F_1$ scores, allowing performance for both proximity classes to be examined in the presence of class imbalance, with higher values meaning better performance. In our evaluation results (see Table~\ref{tab:performance}), class 0 and class 1 cases refer to vessels closer or further than the distance threshold, respectively. 
  \item For the regression task (vessel-to-cable distance estimation): Global Mean Absolute Error (MAE), which is computed considering all the available vessels in the dataset, and the below-$1{,}000\,m$ MAE, computed only for samples whose vessel-to-cable distance is below $1{,}000\,m$, to characterize estimation error in the proximity range most relevant to cable monitoring. Lower values indicate smaller estimation errors.
\end{itemize}

Table~\ref{tab:performance} shows the performance metrics corresponding to $N_C \in \lbrace 10, 50, 250 \rbrace$ channels. To summarize, the classifier achieved a global $F_1$-score of $89.43\%$ when averaging all $250$ channels, with class-wise $F_1$-scores of $83.10\%$ for the vessel-nearby class and $92.47\%$ for the no-vessel-nearby class. For distance estimation, the corresponding global MAE was $829.90\,m$, while the MAE evaluated for samples below $1,000\,m$ was $171.01\,m$. Performance improved consistently as the number of spatial channels used for averaging increased.

\begin{table}[htbp]
  \caption{Performance metrics for the vessel detection and distance estimation tasks under different spatial contexts (\textrm{\# channels} refers to the number of channels considered for averaging).}
  \label{tab:performance}
  \begin{tabular}{ccccccl}
    & \multicolumn{4}{c}{Vessel detection task (distance threshold $1000\,m$)} & \multicolumn{2}{c}{Distance estimation task} \\ \hline
    \# channels & Accuracy     & Global $F_1$     & Class 0 $F_1$     & Class 1 $F_1$     & Global MAE           & Below-$1000\,m$ MAE         \\ \hline
    10          &87.38\%              &87.01\%                   &78.54\%                    &91.06\%                    &$979.84\,m$                      &$196.89\,m$                       \\ \hline
    50          &89.23\%              &88.99\%                   &82.11\%                    &92.29\%                    &$882.79\,m$                      &$191.33\,m$                       \\ \hline
    250         &89.58\%              &89.43\%                   &83.10\%                    &92.47\%                    &$829.90\,m$                      &$171.01\,m$                       \\ \hline
  \end{tabular}
\end{table}

The baseline results show that models trained on the released feature representation can discriminate the defined vessel-proximity classes and estimate vessel-to-cable distance under the reported evaluation protocol. These results provide an application-based validation of the information retained in the processed dataset. Additional experimental analyses using this dataset are reported in~\cite{RamirezTorres2026VesselDAS}.

\section{Usage Notes}
\label{sec:usage notes}

The \texttt{Marlinks-NS DAS} dataset is intended to serve as a benchmark resource for developing and validating ML methods for vessel detection and vessel-to-cable distance estimation using submarine DAS data in submarine cable protection applications. The dataset is structured to enable reproducible experimentation and flexible adaptation to various modeling approaches.

The spectral feature matrices can be directly used as input to ML models for the binary vessel detection or continuous vessel-to-cable distance regression tasks. For custom experiments, users may re-aggregate energy bands, apply normalization, or augment the data using temporal or spatial transformations. The AIS-derived distance labels may also be thresholded to define categorical proximity classes if required, including multi-class classification tasks.

Simple Python examples for basic interaction with the dataset are distributed with the archived Zenodo release in \texttt{src.zip}. These scripts demonstrate how to inspect the HDF5 structure, load the complete feature arrays or selected slices, verify consistency between loading strategies, and generate day-wise ($k$-fold) cross-validation training and test partitions. Further software resources are described in Section~\ref{sec:code availability}.

\section{Data Availability}
\label{sec:data-availability}

The Marlinks-NS DAS dataset and its accompanying documentation are openly available in the Zenodo repository~\cite{ramirez2024dasvesseldataset} at \url{https://doi.org/10.5281/zenodo.15611778}. The versioned deposit contains the complete processed dataset in HDF5 format (\texttt{dataset\_sensor\_range\_1440\_1690\_0.h5}); dataset documentation in Markdown and PDF formats (\texttt{README.md} and \texttt{README.pdf}); supporting citation, licensing, provenance, funding, creators, changelog, DOI, daily ranges of environmental variables, and frequency-band metadata in \texttt{misc.zip}; and simple Python examples for inspecting, loading, checking, and partitioning the HDF5 data in \texttt{src.zip}. The data and accompanying documentation are released under the Creative Commons Attribution 4.0 International license.

\section{Code Availability}
\label{sec:code availability}

A fixed snapshot of the basic Python data-handling scripts associated with the released dataset version is archived in the Zenodo record~\cite{ramirez2024dasvesseldataset} within the \texttt{src.zip} archive.

The \href{https://github.com/UAH-PSI/das-vessel-detection}{companion GitHub repository} provides extended data-handling, partitioning, visualization, and reproducibility tools, and will continue to be actively maintained and updated with corrections, documentation, and new functionality. Because the \href{https://github.com/UAH-PSI/das-vessel-detection}{GitHub repository} will evolve after publication, users should record and report the release, tag, or commit used in their analyses.

The archived scripts require Python 3.8 or later and have been tested under Python 3.10 and 3.11, relying exclusively on widely available open-source Python packages. Dependencies for the extended software are documented in the \href{https://github.com/UAH-PSI/das-vessel-detection}{GitHub repository}. The released source code is distributed under the GNU General Public License v3.0.

Users are encouraged to adapt and extend the provided scripts for their research needs. Contributions to the repository, including improvements, additional examples, and derived analysis tools, are very welcome through standard GitHub pull requests.

\bibliographystyle{IEEEtran-nonote}
\bibliography{paper}

@article{pubDAS2023,
    author = {Spica, Zack J. and Ajo‐Franklin, Jonathan and Beroza, Gregory C. and Biondi, Biondo and Cheng, Feng and Gaite, Beatriz and Luo, Bin and Martin, Eileen and Shen, Junzhu and Thurber, Clifford and Viens, Loïc and Wang, Herbert and Wuestefeld, Andreas and Xiao, Han and Zhu, Tieyuan},
    title = {{PubDAS: A PUBlic Distributed Acoustic Sensing Datasets Repository for Geosciences}},
    journal = {Seismological Research Letters},
    volume = {94},
    number = {2A},
    pages = {983-998},
    year = {2023},
    month = {01},
    issn = {0895-0695},
    doi = {10.1785/0220220279},
    url = {https://doi.org/10.1785/0220220279},
    eprint = {https://pubs.geoscienceworld.org/ssa/srl/article-pdf/94/2A/983/5793670/srl-2022279.1.pdf},
}

@article{tomasov2025comprehensive,
  title={{Comprehensive dataset for event classification using distributed acoustic sensing (DAS) systems}},
  author={Tomasov, Adrian and Zaviska, Pavel and Dejdar, Petr and Klicnik, Ondrej and Horvath, Tomas and Munster, Petr},
  journal={Scientific Data},
  volume={12},
  number={1},
  pages={793},
  year={2025},
  doi={10.1038/s41597-025-05088-4},
  url       = {https://doi.org/10.1038/s41597-025-05088-4},
  publisher={Nature Publishing Group UK London}
}

@Article{Huang2025DAShip,
  author    = {Huang, Wenjin and Chen, Shaoyi and Wu, Yichang and Li, Ruihua and Li, Tianrui and Huang, Yihua and Cao, Xiaochun and Li, Zhaohui},
  title     = {{DAShip: A Large-Scale Annotated Dataset for Ship Detection Using Distributed Acoustic Sensing Technique}},
  journal   = {IEEE Journal of Selected Topics in Applied Earth Observations and Remote Sensing},
  year      = {2025},
  volume    = {18},
  pages     = {4093--4107},
  issn      = {2151-1535},
  doi       = {10.1109/jstars.2024.3525082},
  url       = {https://doi.org/10.1109/jstars.2024.3525082},
  publisher = {Institute of Electrical and Electronics Engineers (IEEE)},
}

@misc{DAShip2025,
  note      = {Dataset},
  author    = {Huang, Wenjin and Chen, Shaoyi and Wu, Yichang and Li, Ruihua and Li, Tianrui and Huang, Yihua and Cao, Xiaochun and Li, Zhaohui},
  title     = {{DAShip dataset}},
  year      = {2025},
  url       = {https://www.alipan.com/s/sTdL3zSRiPo},
}

@misc{gebco2024,
  author = {{GEBCO Bathymetric Compilation Group}},
  title = {{GEBCO 2024 Grid [data set]}},
  year = {2024},
  url = {https://doi.org/10.5285/1c44ce99-0a0d-5f4f-e063-7086abc0ea0f},
  doi = {10.5285/1c44ce99-0a0d-5f4f-e063-7086abc0ea0f},
  OPTpublisher = {British Oceanographic Data Centre, National Oceanography Centre, NERC, UK}
}

@article{cheng2021utilizing,
  title={{Utilizing distributed acoustic sensing and ocean bottom fiber optic cables for submarine structural characterization}},
  author={Cheng, Feng and Chi, Benxin and Lindsey, Nathaniel J and Dawe, T Craig and Ajo-Franklin, Jonathan B},
  journal={Scientific reports},
  volume={11},
  number={1},
  pages={5613},
  year={2021},
  doi={10.1038/s41598-021-84845-y},
  url       = {https://doi.org/10.1038/s41598-021-84845-y},
  publisher={Nature Publishing Group UK London}
}

@misc{cheng2020mbari_das_data,
  note      = {Dataset: Utilizing Distributed Acoustic Sensing and Ocean Bottom Fiber Optic Cables for Submarine Characterization},
  title        = {{Dataset from the mBARI DAS Project}},
  author       = {Cheng, Feng},
  year         = {2020},
  howpublished = {Dataset},
  OPTinstitution  = {--},
  doi          = {10.17605/OSF.IO/CN8XB},
  url          = {https://doi.org/10.17605/OSF.IO/CN8XB},
  date_created = {2020-10-20},
  date_updated = {2025-02-21},
  doi          = {10.17605/OSF.IO/CN8XB}
}

@article{lior2021detection,
  title={On the detection capabilities of underwater distributed acoustic sensing},
  author={Lior, Itzhak and Sladen, Anthony and Rivet, Diane and Ampuero, Jean-Paul and Hello, Yann and Becerril, Carlos and Martins, Hugo F and Lamare, Patrick and Jestin, Camille and Tsagkli, Stavroula and others},
  journal={Journal of Geophysical Research: Solid Earth},
  volume={126},
  number={3},
  pages={e2020JB020925},
  year={2021},
  doi={10.1029/2020JB020925},
  url       = {https://doi.org/10.1029/2020JB020925},
  publisher={Wiley Online Library}
}

@misc{lior2021detection_dataset,
  title        = {{The Underwater DAS Detection Dataset}},
  author       = {Itzhak Lior},
  year         = {2019},
  note         = {Dataset: Distributed Acoustic Sensing experiment done on the MEUST-NUMerEnv/KM3NeT fiber optic cable, laid offshore Toulon, south of France. Data associated to the work presented in the manuscript available at \url{https://eartharxiv.org/ekrfy} and entitled: \emph{Distributed sensing of earthquakes and ocean-solid Earth interactions on seafloor telecom cables}. Authors: Anthony Sladen, Diane Rivet, Jean-Paul Ampuero, Louis De Barros, Yann Hello, Gaëtan Calbris, Patrick Lamare.},
  howpublished = {Dataset},
  date         = {2019-07-29},
  urldate      = {2019-12-06},
  doi          = {10.17605/OSF.IO/4BJPH},
  url       = {https://doi.org/10.17605/OSF.IO/4BJPH}
}

@article{sladen2019distributed,
  title={{Distributed sensing of earthquakes and ocean-solid Earth interactions on seafloor telecom cables}},
  author={Sladen, Anthony and Rivet, Diane and Ampuero, Jean Paul and De Barros, Louis and Hello, Yann and Calbris, Ga{\"e}tan and Lamare, Patrick},
  journal={Nature communications},
  volume={10},
  number={1},
  pages={5777},
  year={2019},
  doi={10.1038/s41467-019-13793-z},
  url       = {https://doi.org/10.1038/s41467-019-13793-z},
  publisher={Nature Publishing Group UK London}
}

@misc{sladen2019meust_km3net_das,
  title        = {{Dataset for the MEUST-NUMerEnv/KM3NeT DAS experiment Feb. 2018}},
  author       = {Anthony Sladen},
  year         = {2019},
  note         = {Dataset: Distributed Acoustic Sensing experiment done on the MEUST-NUMerEnv/KM3NeT fiber optic cable, laid offshore Toulon, south of France. Data associated to the work presented in the manuscript available at \url{https://eartharxiv.org/ekrfy} and entitled: \emph{Distributed sensing of earthquakes and ocean-solid Earth interactions on seafloor telecom cables}. Authors: Anthony Sladen, Diane Rivet, Jean-Paul Ampuero, Louis De Barros, Yann Hello, Gaëtan Calbris, Patrick Lamare.},
  howpublished = {Dataset},
  date         = {2019-07-29},
  urldate      = {2019-12-06},
  doi          = {10.17605/OSF.IO/X6AW},
  url       = {https://doi.org/10.17605/OSF.IO/X6AWB}
}

@Article{Taweesintananon2023,
  author    = {Taweesintananon, Kittinat and Landrø, Martin and Potter, John Robert and Johansen, Ståle Emil and Rørstadbotnen, Robin André and Bouffaut, Léa and Kriesell, Hannah Joy and Brenne, Jan Kristoffer and Haukanes, Aksel and Schjelderup, Olaf and Storvik, Frode},
  title     = {{Distributed acoustic sensing of ocean-bottom seismo-acoustics and distant storms: A case study from Svalbard, Norway}},
  journal   = {GEOPHYSICS},
  year      = {2023},
  volume    = {88},
  number    = {3},
  pages     = {B135–B150},
  OPTmonth     = apr,
  issn      = {1942-2156},
  doi       = {10.1190/geo2022-0435.1},
  url       = {https://doi.org/10.1190/geo2022-0435.1},
  publisher = {Society of Exploration Geophysicists},
}

@misc{VPRD2H_2022,
  note      = {Dataset},
author = {Taweesintananon, Kittinat and Landrø, Martin},
publisher = {DataverseNO},
title = {{Replication data for DAS4Microseism - Svalbard distributed acoustic sensing (DAS) strain data for oceanographic study}},
year = {2022},
version = {V1},
doi = {10.18710/VPRD2H},
url = {https://doi.org/10.18710/VPRD2H}
}

@article{shi2025multiplexed,
  title={{Multiplexed distributed acoustic sensing offshore central Oregon}},
  author={Shi, Qibin and Williams, Ethan F and Lipovsky, Bradley P and Denolle, Marine A and Wilcock, William SD and Kelley, Deborah S and Schoedl, Katelyn},
  journal={Seismological Research Letters},
  volume={96},
  number={2A},
  pages={784--800},
  year={2025},
  doi={10.1785/0220240460},
  url       = {https://doi.org/10.1785/0220240460},
  publisher={Seismological Society of America}
}

@misc{lipovsky2024rapid_das_ooi,
  note      = {Dataset},
  author       = {Lipovsky, Bradley and Williams, Evan and Ocean Observatories Initiative},
  title        = {{{RAPID}: Multiplexed Distributed Acoustic Sensing ({DAS}) at the Ocean Observatory Initiative ({OOI}) Regional Cabled Array ({RCA})}},
  year         = {2024},
  publisher    = {Ocean Observatories Initiative},
  doi          = {10.58046/4WEF-A282},
  url          = {https://doi.org/10.58046/4WEF-A282},
  OPTnote         = {Data set licensed under Creative Commons Attribution-NonCommercial-ShareAlike 4.0 International. NSF Award \#2415521: RAPID: Multiplexed Distributed Acoustic Sensing (DAS) at the Ocean Observatory Initiative (OOI) Regional Cabled Array (RCA).},
  howpublished = {Dataset}
}

@article{song2024near,
  title={Near real-time in situ monitoring of nearshore ocean currents using distributed acoustic sensing on submarine fiber-optic cable},
  author={Song, Zhenghong and Zeng, Xiangfang and Ni, Sidao and Chi, Benxin and Xu, Tengfei and Wei, Zexun and Jiang, Wenzheng and Chen, Sheng and Xie, Jun},
  journal={Earth and Space Science},
  volume={11},
  number={9},
  pages={e2024EA003572},
  year={2024},
  doi={10.1029/2024EA003572},
  url       = {https://doi.org/10.1029/2024EA003572},
  publisher={Wiley Online Library}
}

@misc{song_2024_dataset,
note = {Dataset},
  author       = {Song, Zhenghong and
                  Zeng, Xiangfang and
                  Ni, Sidao and
                  Chi, Benxin and
                  Xu, Tengfei and
                  Wei, Zexun and
                  Jiang, Wenzheng and
                  Chen, Sheng and
                  Xie, Jun},
  title        = {{Dataset in ``Near real-time in-situ monitoring of
                   nearshore ocean currents using Distributed
                   Acoustic Sensing on submarine fiber-optic cable''
                  }},
  OPTmonth        = jul,
  year         = 2024,
  publisher    = {Zenodo},
  doi          = {10.5281/zenodo.13133835},
  url          = {https://doi.org/10.5281/zenodo.13133835},
}

@Article{Rivet_2021,
  author    = {Rivet, Diane and de Cacqueray, Benoit and Sladen, Anthony and Roques, Aurélien and Calbris, Gaëtan},
  title     = {{Preliminary assessment of ship detection and trajectory evaluation using distributed acoustic sensing on an optical fiber telecom cable}},
  journal   = {The Journal of the Acoustical Society of America},
  year      = {2021},
  volume    = {149},
  number    = {4},
  pages     = {2615--2627},
  OPTmonth     = apr,
  issn      = {1520-8524},
  note      = {Abstract Distributed acoustic sensing (DAS) is a recent instrumental approach allowing the conversion of fiber-optic cables into dense arrays of acoustic sensors. This technology is attractive in marine environments where instrumentation is difficult to implement. A promising application is the monitoring of environmental and anthropic noise, leveraging existing telecommunication cables on the seafloor. We assess the ability of DAS to monitor such noise using a 41.5 km-long cable offshore of Toulon, France, focusing on a known and localized source. We analyze the noise emitted by the same tanker cruising above the cable, first 5.8 km offshore in 85 m deep bathymetry, and then 20 km offshore, where the seafloor is at a depth of 2000 m. The spectral analysis, the Doppler shift, and the apparent velocity of the acoustic waves striking the fiber allow us to separate the ship radiated noise from other noise. At 85 m water depth, the signal-to-noise ratio is high, and the trajectory of the boat is recovered with beamforming analysis. At 2000 m water depth, although the acoustic signal of the ship is more attenuated, signals below 50 Hz are detected. These results confirm the potential of DAS applied to seafloor cables for remote monitoring of acoustic noise even at intermediate depth.},
  doi       = {10.1121/10.0004129},
  url       = {https://doi.org/10.1121/10.0004129},
  publisher = {Acoustical Society of America (ASA)},
}

@InProceedings{Thiem_2023,
  author    = {Thiem, Lukas and Wienecke, Susann and Taweesintananon, Kittinat and Vaupel, Melvin and Landrø, Martin},
  title     = {{Ship noise characterization for marine traffic monitoring using distributed acoustic sensing}},
  booktitle = {{2023 IEEE International Workshop on Metrology for the Sea; Learning to Measure Sea Health Parameters}},
  year      = {2023},
  pages     = {334--339},
  OPTmonth     = oct,
  publisher = {IEEE},
  note      = {Ship_noise_characterization_for_marine_traffic_monitoring_using_distributed_acoustic_sensing.pdf Typical Sound Frequencies of Different Ship Types: The study uses band-pass filtering from 100–120 Hz to focus on vessel-generated P-wave signals, emphasizing this range for noise reduction and analysis in shallow, noisy environments. Background on Ship Detection Using Machine Learning: While not directly applying ML, the study integrates signal processing and novel methodologies (e.g., persistent homology) to detect and track vessel signals. These approaches provide a foundation for ML-based feature engineering and classification models. Machine Learning Applied to Submarine Optical Fiber Signals: Persistent homology and image processing techniques, such as Gaussian filtering and Otsu’s thresholding, are employed to enhance signal detection. These techniques can be adapted for preprocessing and feature extraction in ML models using DAS data. Other Useful Highlights: Demonstrates the feasibility of DAS for marine traffic monitoring, even with poorly coupled cables in soft sediments. Successfully localizes vessels using direct P-wave arrivals and travel-time inversion, achieving results comparable to GPS tracking from AIS. Highlights the potential of DAS for detecting "dark ships" (vessels without AIS signals), which is significant for maritime security and environmental monitoring. Discusses the challenges and limitations of shallow water environments, such as uncertainties in bathymetry, sound velocity, and cable positioning, which affect localization accuracy. Provides a roadmap for integrating DAS with automatic algorithms and edge computing for future ML applications in signal classification.},
  doi       = {10.1109/metrosea58055.2023.10317227},
  url       = {https://doi.org/10.1109/metrosea58055.2023.10317227}
}

@misc{navalnews_estlink2_2024,
  title = {{Seabed Cable Damaged in Latest Baltic CUI Incident}},
  author = {{Naval News}},
  year = {2024},
  OPTurl = {https://www.navalnews.com/naval-news/2024/12/seabed-cable-damaged-in-latest-baltic-cui-incident/},
  howpublished = {\href{https://www.navalnews.com/naval-news/2024/12/seabed-cable-damaged-in-latest-baltic-cui-incident/}{Online, accessed July 2026}},
}

@misc{euromaidan_baltic_2025,
  title = {{Finland-Germany submarine cable damaged again in Baltic Sea in possible sabotage act}},
  author = {Yuri Zoria},
  year = {2025},
  OPTurl = {https://euromaidanpress.com/2025/02/21/finland-germany-submarine-cable-damaged-again-in-baltic-sea-in-possible-sabotage-act/},
  howpublished = {\href{https://euromaidanpress.com/2025/02/21/finland-germany-submarine-cable-damaged-again-in-baltic-sea-in-possible-sabotage-act/}{Online, accessed July 2026}},
}

@article{RamirezTorres2026VesselDAS,
  author  = {Ramirez-Torres, Erick Eduardo and Macias-Guarasa, Javier and Pizarro-Perez, Daniel and Tejedor, Javier and Palazuelos-Cagigas, Sira Elena and Vidal-Moreno, Pedro J. and Martin-Lopez, Sonia and Gonzalez-Herraez, Miguel and Vanthillo, Roel},
  title   = {Vessel Detection and Localization Using Distributed Acoustic Sensing in Submarine Optical Fiber Cables (accepted for publication)},
  journal = {IEEE Journal of Selected Topics in Applied Earth Observations and Remote Sensing},
  year    = {2026},
  OPTvolume  = {XX},
  OPTnumber  = {XX},
  OPTpages   = {XX--XX},
  note    = {In press},
  doi     = {10.1109/JSTARS.2026.3716768},
  url     = {https://doi.org/10.1109/JSTARS.2026.3716768},
}

@misc{OptoDAS,
author = {{Alcatel Submarine Networks}},
OPTtitle = {\href{https://www.asn.com/fiber-sensing/}{OptoDAS interrogator}},
title = {{OptoDAS interrogator}},
  OPThowpublished = {Online, accessed July 2026},
  howpublished = {\href{https://www.asn.com/fiber-sensing/}{Online, accessed July 2026}},
  OPTurl = {https://www.asn.com/fiber-sensing/},
  note         = {}
}

@article{gabai2016sensitivity,
  title={{On the sensitivity of distributed acoustic sensing}},
  author={Gabai, Haniel and Eyal, Avishay},
  journal={Optics letters},
  volume={41},
  number={24},
  pages={5648--5651},
  year={2016},
  doi = {10.1364/OL.41.005648},
  url       = {https://doi.org/10.1364/OL.41.005648},
  publisher={Optical Society of America}
}

@Article{Zou_2015,
  author    = {Zou, Weiwen and Yang, Shuo and Long, Xin and Chen, Jianping},
  title     = {{Optical pulse compression reflectometry: proposal and proof-of-concept experiment}},
  journal   = {Optics Express},
  year      = {2015},
  volume    = {23},
  number    = {1},
  pages     = {512},
  OPTmonth     = jan,
  issn      = {1094-4087},
  doi       = {10.1364/oe.23.000512},
  url       = {https://doi.org/10.1364/oe.23.000512},
  publisher = {Optica Publishing Group},
}

@InProceedings{Waagaard_2021,
  author     = {Waagaard, Ole Henrik and Rønnekleiv, Erlend and Haukanes, Aksel and Stabo-Eeg, Frantz and Thingbø, Dag and Forbord, Stig and Aasen, Svein Erik and Brenne, Jan Kristoffer},
  title      = {{Real-time phase-recording DAS in 171 km low-loss fiber}},
  booktitle  = {{Optical Fiber Sensors Conference 2020 Special Edition}},
  year       = {2021},
  series     = {OFS},
  pages      = {T2A.3},
  publisher  = {Optica Publishing Group},
  OPTcollection = {OFS},
  doi        = {10.1364/ofs.2020.t2a.3},
  url       = {https://doi.org/10.1364/ofs.2020.t2a.3}
}

@misc{ramirez2024dasvesseldataset,
note = {Dataset},
  author       = {Erick Eduardo Ramirez-Torres and
                  Javier Macias-Guarasa and
                  Daniel Pizarro-Perez and
                  Javier Tejedor and
                  Sira Elena Palazuelos-Cagigas and
                  Pedro Vidal-Moreno and
                  Marı́a R. Fernández-Ruiz and
                  Sonia Martin-Lopez and
                  Miguel Gonzalez-Herraez and
                  Roel Vanthillo},
  title        = {{Marlinks-NS DAS: Dataset for Vessel Detection and Distance Estimation Using Distributed Acoustic Sensing in Submarine Cables}},
  year         = {2026},
  publisher    = {Zenodo},
  doi          = {10.5281/zenodo.15611778},
  url          = {https://doi.org/10.5281/zenodo.15611778}
}

@Article{holman2025,
AUTHOR = {Holman, Robert and Glover, Hannah and Wengrove, Meagan and Ifju, Marcela and Honegger, David and Haller, Merrick},
TITLE = {{Geolocation of Distributed Acoustic Sampling Channels Using X-Band Radar and Optical Remote Sensing}},
JOURNAL = {Remote Sensing},
VOLUME = {17},
YEAR = {2025},
NUMBER = {18},
ARTICLE-NUMBER = {3142},
OPTURL = {https://www.mdpi.com/2072-4292/17/18/3142},
ISSN = {2072-4292},
DOI = {10.3390/rs17183142},
pages={3142},
url       = {https://doi.org/10.3390/rs17183142}
}

@Article{emmens2021promisesAIS,
  author    = {Emmens, Ties and Amrit, Chintan and Abdi, Asad and Ghosh, Mayukh},
  title     = {{The promises and perils of Automatic Identification System data}},
  journal   = {Expert Systems with Applications},
  year      = {2021},
  volume    = {178},
  pages     = {114975},
  doi={10.1016/j.eswa.2021.114975},
  url       = {https://doi.org/10.1016/j.eswa.2021.114975},
  note      = {1-s2.0-S0957417421004164-main.pdf AIS Data Usage in Maritime Applications: AIS (Automatic Identification System) is crucial for identifying vessels in navigation, environmental monitoring, and improving navigational safety. Applications include route optimization, collision prevention, and emissions estimation. Challenges with AIS Data: Noise: Significant noise is present in static, dynamic, and voyage-related AIS data. For instance, positional data and Speed over Ground (SOG) often exhibit inconsistencies. Equipment Quality: Variability in equipment performance impacts data accuracy. Issues include unstable Course over Ground and incorrect SOG values. Human Factors: Errors stem from incorrect data entry, equipment misconfiguration, and deliberate signal switch-offs. Applications Highlighted: Safety and Collision Avoidance: AIS supports navigational safety by enabling vessel identification and communication. Environmental Monitoring: AIS can assess adherence to environmental regulations, such as restricted areas or emissions tracking. Supply Chain Optimization: Predictive analytics using AIS enhances port management and facilitates automatic tax collection. Research Insights: Mixed-method design combining quantitative AIS data analysis (from the Port of Amsterdam) with qualitative expert interviews provides validation. Identifies systemic issues like incomplete tracks, data gaps, and the need for better integration of AIS with other systems (e.g., radar, geographic data). Future Directions: Addressing equipment inconsistencies and integrating advanced data filtering to mitigate noise. Enhancing the reliability of static and voyage-related AIS data for broader applications.},
  publisher = {Elsevier},
}

@article{guo2021improved,
  title={{Improved kinematic interpolation for AIS trajectory reconstruction}},
  author={Guo, Shaoqing and Mou, Junmin and Chen, Linying and Chen, Pengfei},
  journal={Ocean Engineering},
  volume={234},
  pages={109256},
  year={2021},
  doi={10.1016/j.oceaneng.2021.109256},
  url       = {https://doi.org/10.1016/j.oceaneng.2021.109256},
  publisher={Elsevier}
}


\section*{Author Contributions}

All authors defined the released dataset, conceived the study, and designed the general experimental methodology and machine-learning tasks. R.V. coordinated the acquisition of and access to the DAS data, acquisition metadata, and preprocessing software. P.J.V.-M, M.R.F.-R., S.M.-L. and M.G.-H. contributed to DAS preprocessing and optical-signal interpretation. E.E.R.-T., J.M.-G., D.P.-P. and S.E.P.-C. processed the AIS and geospatial information, synchronized it with the DAS measurements, and generated the vessel-distance labels. E.E.R.-T., D.P.-P. and J.M.-G. performed feature extraction, dataset curation, and design of the released HDF5 structure. E.E.R.-T., J.M.-G., D.P.-P., J.T. and S.E.P.-C. developed the machine-learning software and carried out the technical validation. E.E.R.-T., J.M.-G. and S.E.P.-C. prepared and published the data and code repositories, and drafted the manuscript. J.M.-G., M.R.F.-R., S.M.-L. and M.G.-H. provided scientific supervision, project coordination, and funding acquisition. All authors reviewed and approved the manuscript.

\section*{Competing Interests}
\label{sec:competing interests}

The authors declare no competing interests.

\section*{Acknowledgments}
\label{sec:acknowledgements}

We gratefully acknowledge the computer resources at Artemisa, funded by the ``European Union ERDF'' and ``Comunitat Valenciana'' as well as the technical support provided by the ``Instituto de Física Corpuscular'', IFIC (CSIC-UV). We also thank the cable monitoring operator and the cable owner for allowing data access under confidentiality requirements.

\section*{Funding}
\label{sec:funding}

This work has been partially supported by the ``Spanish Ministry of Science and Innovation'' MICIU/AEI/10.13039/501100011033, FEDER UE, and by the ``European Union NextGeneration EU/PRTR'' program under grants PSI (PLEC2021-007875), NeurEYE-UAH (PID2024-156576OB-C31), SEASNAKE+ (PCI2023-145978-2, of the CETPartnership 2022 joint call), and MOTION (PID2022-140963OA-I00); by the ``European Innovation Council'' under grant ECSTATIC (101189595);  and by the European Research Council under grant SENSE (101218803). The work of M.R.F-R. was also supported by MCIN/AEI/10.13039/501100011033 and European Union ``NextGenerationEU/PRTR'' program under grant RYC2021-032167-I.


\end{document}